\documentclass[journal]{IEEEtran}
\ifCLASSINFOpdf
\else
\fi

\usepackage{mathrsfs}
\usepackage{color}
\usepackage{amsfonts,amssymb}
\usepackage{amsfonts}
\usepackage{subfigure}
\usepackage{booktabs}

\usepackage{amsmath}
\usepackage{enumerate}
\usepackage{algorithm}
\usepackage{algorithmic}
\usepackage{bm}
\usepackage{makecell}

\usepackage{mathtools}
\usepackage{bbm}
\usepackage{dsfont}
\usepackage{pifont}
\usepackage{amsthm}
\usepackage{cite}

\newtheorem{remark}{Remark}
\newtheorem{theorem}{Theorem}
\newtheorem{lemma}{Lemma}
\newtheorem{corollary}{Corollary}

\newtheorem{definition}{Definition}

\let \sss=\scriptscriptstyle

\begin{document}

\title{
Sneak Attack against Mobile Robotic Networks under Formation Control}
\author{Yushan Li$^{\dag}$, Jianping He$^{\dag}$, Xuda Ding$^{\dag}$, Lin Cai$^{\ddag}$ and Xinping Guan$^{\dag}$
	\thanks{
	${\dag}$: The Dept. of Automation, Shanghai Jiao Tong University, and Key Laboratory of System Control and Information Processing, Ministry of Education of China, Shanghai, China. E-mail address: \{yushan\_li, jphe, dingxuda, xpguan\}@sjtu.edu.cn. 

	${\ddag}$: The Dept. of Electrical and Computer Engineering, University of Victoria, BC, Canada. Email address: cai@ece.uvic.ca.

	Preliminary results have been submitted to 2021 European Control Conference \cite{lys}.
	}%
}

\maketitle


\begin{abstract}
The security of mobile robotic networks (MRNs) has been an active research topic in recent years. 
This paper demonstrates that the observable interaction process of MRNs under formation control will present increasingly severe threats. 
Specifically, we find that an external attack robot, who has only partial observation over MRNs while not knowing the system dynamics or access, can learn the interaction rules from observations and utilize them to replace a target robot, destroying the cooperation performance of MRNs. 
We call this novel attack as sneak, which endows the attacker with the intelligence of learning knowledge and is hard to be tackled by traditional defense techniques. 
The key insight is to separately reveal the internal interaction structure within robots and the external interaction mechanism with the environment, from the coupled state evolution influenced by the model-unknown rules and unobservable part of the MRN. 
To address this issue, we first provide general interaction process modeling and prove the learnability of the interaction rules. 
Then, with the learned rules, we design an Evaluate-Cut-Restore (ECR) attack strategy considering the partial interaction structure and geometric pattern. 
We also establish the sufficient conditions for a successful sneak with maximum control impacts over the MRN. 
Extensive simulations illustrate the feasibility and effectiveness of the proposed attack.

\end{abstract}

\section{Introduction}
Mobile robotic networks (MRNs) have received considerable attention in recent years. 
Thanks to the mobility, flexibility, and distributed fashion, MRNs are widely deployed in numerous applications, e.g., surveillance, reconnaissance, search and environmental monitoring \cite{bullo2009distributed}. 
Among these applications, formation control is a fundamental technique to enhance the cooperation performance by maintaining a preset geometric shape \cite{olfati2004consensus,oh2015survey,10.1145/3366696,10.1145/3366701,kamel2020formation}.

It has been reported that MRNs under formation control may suffer severe security vulnerabilities from both cyber and physical spaces \cite{choi2018detecting,10.1145/3366698}. 
Existing works mainly focus on the cybersecurity of MRNs, and various defense techniques are developed to protect the information availability, integrity and confidentiality. 
In the literature, the attacker is generally assumed with strong knowledge about the system (e.g., network structure, dynamics model) or powerful capabilities (e.g., access to the system), lacking for deep investigation on the capabilities and modes of attackers in more realistic situations. 
This paper demonstrates that an external attacker without strong prior knowledge can observe the state evolution of the MRN to obtain data, learn the interaction rules from the data analysis, and further leverage the rules to implement attacks that cannot be handled by traditional security countermeasures. 
Several prior works have made some progress in this direction and support our demonstration. 
For example, it is possible to infer whether two robots are connected \cite{8985069} or learn how a single robot avoids an obstacle \cite{li2019learning}. 
However, considering what exact interaction rules can be learned and under what conditions to obtain them, how to further destroy the cooperation performance by leveraging the rules and where is the attack impact boundary, they are all still open issues and need further exploration. 

Motivated by above discussions, we design a sneak attack against MRNs under formation control, where the attack robot has only partial observation and aims to replace one robot therein, obtaining partial control over the MRN and destroying the cooperation performance. 
This attack is inspired by that i) the shape-forming process essentially reflects the internal interaction structure among cooperating robots \cite{olfati2007consensus}; 
ii) the obstacle-avoidance behavior embodies how the robots interact with the environment \cite{pandey2017mobile}; 
iii) both the interaction rules are limited by physical distance in practical applications, either by communication or sensing. 
These universal interaction characteristics indicate the possibility of learning them to implement advanced attacks, making our work of both theoretical and practical significance. 
From the attack view, it enlightens the feasibility of revealing the interaction rules that support the MRN, and guides for more advanced attack design (e.g., invade an MRN). 
In turn, from the defense view, it provides deeper insights into the trade-off between the system security and observability, and helps the countermeasure design by understanding the possible attack capabilities and modes.

\textbf{Challenges and contributions}. 
The major challenges of achieving the sneak attack lie in two aspects. 
First, the attacker can only observe the measurable states of a partial MRN (e.g., position, velocity), and the states are influenced by both the model-unknown interaction rules and the unobservable part of the MRN. 
It is difficult to split the coupled influences of the multiple factors and approximate the interaction rules. 
Second, even if the interaction rules are obtained, their mutual impacts will closely intersect with the influence of the attack dynamics, making it further hard to design feasible and efficient strategies for the attacker to achieve maximum attack rewards. 
To overcome above issues, we holistically formulate the entire observe (excite)-attack process as different stages. 
By exploiting the independence and correlation characteristics of the interaction rules in different stages, the whole attack mode is designed in an active-knowledge-growth form and accommodates the coupled dynamics of the attacker and the MRN.

The differences between this paper and its conference version \cite{lys} include i) a systematic intelligent attack design is introduced rather than mere topology inference, 
ii) the learnability of both the internal interaction structure within the MRN and the external interaction mechanism with the environment is explored, 
and iii) the sneak attack is introduced with comprehensive feasibility and performance analysis. 
The main contributions of this paper are summarized as follows. 

\begin{itemize}
\item We contribute to the existing body of research by understanding what interaction rules in MRNs under formation can be separately revealed by observing over the state evolution, which is a mutual consequence of multiple factors. 
No strong prior information like dynamics model or access to the MRN is required. 

\item We find a novel sneak attack where an attacker with weak knowledge can learn the interaction rules and replace a victim robot to obtain partial control over the MRN. 
The observation and excitation oriented ideas endow the ``ignorant'' attacker with growing intelligence to achieve maximum attack impacts, and this kind of attack is hard to be tackled by traditional defense techniques. 

\item We develop an in-depth framework to handle different interaction rules of MRNs. 
We prove the learnability of the interaction rules by extracting their universal characteristics. 
Utilizing the proposed node controllability, we design an ECR strategy and establish the sufficient conditions for the attack. 
Performance study by simulation demonstrates the effectiveness of the sneak attack. 
\end{itemize}

In a larger sense, this paper promotes to explore the security vulnerabilities in the interaction process, and beckons further research to address the new physical observation and excitation oriented attacks on generic networked systems. 

\textbf{Organization and Notation}. 
The remainder of this paper is organized as follows. 
Section \ref{r-work} presents related literatures. 
Section \ref{preliminary} gives the modeling for the interaction process of MRNs and formulates the attack process. 
Section \ref{revealing} studies how to reveal the interaction rules. 
Section \ref{attack} develops how to design control strategies for the final sneak. 
Simulation results are shown in Section \ref{simulation}, 
followed by the concluding remarks and further research issues in Section \ref{conclusion}. 
All the proofs of the Lemmas and Theorems are provided in the Appendix. 

Throughout the paper, the set variable is expressed in Euclid font, and the compact all-dimension state of a robot is expressed in bold font. 
Unless otherwise noted, the formulas with non-bold font state variable apply to every dimension of the actual state space. 
$\mathcal{A} \backslash \mathcal{B}$ represents the elements in $\mathcal{A}$ that are not in $\mathcal{B}$. 
The superscripts ${[i,:]}$, ${[:,i]}$ of a matrix denote its $i$-th row vector and $i$-th column vector, respectively. 
The scripts $\tilde \cdot $ and  $\hat \cdot$ right above a variable indicate its corresponding observation and estimation value, respectively. 
The basic notation definitions are given in Table \ref{tab:test}.

\section{Related Work}\label{r-work}
\textit{Formation control in MRNs}. 
The fundamental rules for formation control were first introduced by the famous Reynolds' Rules \cite{reynolds1987flocks}: separation, alignment, and cohesion. 
Based on the rules, numerous methods have been proposed to achieve the desired performance, and consensus-based algorithms become the mainstream, e.g., \cite{sun2016optimal,zhao2018affine,alonso2019distributed,xu2020affine}. 
The key idea of consensus-based algorithms is that the formation is modeled as a graph, and every robot implements information interaction (like positions and velocities) with their neighbors and compute their control inputs. 
Therefore, the interaction structure lays critical support for effective formation control and is mostly based on network communication. 
In recent years, communication-free formation control \cite{deghat2014localization,cheng2017event,trinh2018bearing} was also developed and attracts great research interests, thanks to the emergent development of sensor technology. 
Communication-free interaction avoids information delays and network bandwidth consumption, and even enables stealth modes of operation \cite{kan2011network}. 
For instance, formation control with only bearing measurements by vision sensors was investigated in \cite{zhao2019bearing}. 

Although the Reynolds' Rules lay a solid foundation for formation control, they do not take into account some unanticipated faults (e.g., interaction failure, formation splitting), which are easily incurred when performing tasks in the real environment. 
Many advanced interaction rules and techniques were developed to better handle those accidental and unpremeditated faults \cite{feng2016robust,khalili2018distributed,liu2018hierarchical}. 
For instance, a consensus-based splitting/merging algorithm was designed to tackle complex obstacle distributions in \cite{zhu2019distributed}. 
A topology recovery method was designed to reconnect a broken link based on historical data in \cite{li2019optimal}. 
\cite{Zhe2015A,liu2017self} proposed a gradient-based self-repairing algorithm to restore the topology of formation and further synchronize the formation motion. 
The motion synchronization methods were investigated to form desired/predefined formation configuration in the presence of robot malfunctions \cite{liu2018failure} or undesirable formation variation \cite{dong2017time}. 
\cite{yang2019fault} gave a comprehensive survey for more types of fault-tolerant design about formation control. 
Note that by either way, the interaction capability is limited by distance due to the energy constraint in real-world applications, and this simple yet crucial point is utilized in our attack design.

\textit{The security of MRNs}. 
Effective formation control in MRNs is supported by physical sense and network interaction, which constitute the feedback loops for control and also incur security vulnerabilities, and an attack on these components may cause disastrous consequences. 
MRN under formation control a typical cyber-physical-system (CPS), while the existing work mainly focused on the defense design to cyber attacks, e.g., DoS, false data injection, and replay attacks \cite{pasqualetti2013attack,mo2015performance,sandberg2015cyberphysical,sanchez2019bibliographical}. 
The attacker is generally assumed with strong prior knowledge about the MRN (like knowing the system structure or node states \cite{pasqualetti2012consensus}, corrupting exchanged information \cite{silvestre2017stochastic}), and formulated with a powerful attack model. 
Recently, a few works emerged which are dedicated to attacks against formation from physical space. 
For instance, a noise-generating attack was proposed in \cite{son2015rocking} to alter gyroscopic sensor data, leading to drone crashes. 
\cite{tippenhauer2011requirements,bianchin2019secure} proposed a spoofing attack to disturb the GPS sensor readings, gradually causing trajectory deviations. 
These attacks are against a specific transducer by utilizing its sensing mechanism, and are hard to be generalized to other scenarios. 
\cite{li2019learning} developed a trial-and-learning based method to infer the obstacle-avoidance mechanism of mobile robots by disguising the attack robot as an obstacle, which requires little prior knowledge about mobile robots. 

In summary, concerning what attack capability is feasible for an ``ignorant'' attacker from external observation and what is the upper limit of the attack impacts on MRNs, it still remains an open yet critical issue that motivates this work.

\section{Preliminaries and Problem Formulation}\label{preliminary}
\subsection{Graph Basics}
Let $\mathcal{G}=(\mathcal{V},\mathcal{E})$ be a directed graph that models an MRN, where $\mathcal{V}=\{1, \cdots, {N}\}$ is the finite set of nodes (i.e., robots) and $\mathcal{E}\subseteq \mathcal{V}\times \mathcal{V}$ is the set of interaction edges. 
An edge $(i,j)\in \mathcal{E}$ indicates that $i$ will use information from $j$. 
The adjacency matrix $A=[a_{ij}]_{N \times N}$ of $\mathcal{G}$ is defined such that ${a}_{ij}\!>\!0$ if $(i,j)$ exists, and ${a}_{ij}\!=\!0$ otherwise. 
Denote ${\mathcal{N}_i^{in}}=\{j\in \mathcal{V}:a_{ij}>0\}$ and ${\mathcal{N}_i^{out}}=\{j\in \mathcal{V}:a_{ji}>0\}$ as the in-neighbor and  out-neighbor sets of $i$, respectively. 
A directed path is a sequence of nodes $\left \{1, 2, \cdots, j\right\}$ such that $({i+1},i)\!\in\!\mathcal{E}$, $i=1, 2, \cdots, j-1$. 
A directed graph has a (directed) spanning tree if there exists at least a node having a directed path to all other nodes. 
$\mathcal{G}$ must have a spanning tree to guarantee the information flow in robot formation.

Define $L=\text{diag}\{A \bm{1}\}-A$ as the Laplacian matrix of $\mathcal{G}$, where $\bm{1}$ is a vector of all ones. 
Then, $L\bm{1}=0$ holds. 
Let $L=\Gamma \Lambda \Gamma^{-1}$ be the Jordan decomposition of $L$, where $\Gamma=[q_1 \cdots q_{\sss N}]$ is the transformation matrix, $\Gamma^{-1}=[p_1 \cdots p_{\sss N}]$, and $\Lambda  \!=\! \text{diag}\{\bm{\lambda_i}\}$ with $\bm{\lambda_i}$ being the Jordan block of eigenvalue $\lambda_i$.
The eigenvalues and eigenvectors satisfy
\begin{equation}\label{f1}
({\lambda _i}I - L){q_i} = 0,~p_i^\mathsf{T}({\lambda _i}I - L) = 0,~p_i^\mathsf{T}q_j\!=\!0~(i\!\ne \!j).
\end{equation}
Specially, take $q_1=\bm{1}$ and ${p_1 = {[{p_{\sss 11}} \cdots {p_{\sss 1N}}]^\mathsf{T}}}$. 
All the right and left eigenvector are normalized such that ${p_i^\mathsf{T}}{q_i} = 1$.

\begin{table}[t]
\small
 \caption{\label{tab:test}Notation Definitions} 
 \begin{tabular}{cl}
 \toprule 
  Symbol  & Definition  \\ 
  \midrule
   $z_i$ & the state of robot $i$ ($r_i$) in one dimension\\
  $z_a$ & the state of attack robot ($r_a$) in one dimension\\
  $z_v$ & the state of victim robot ($r_v$) in one dimension\\
  $\bm{z}_{\omega}$ & the compact state of a robot in all dimensions\\
  $g$ & the external interaction mechanism with environment\\
  $W$ & the internal interaction matrix among robots\\
  $R_{\omega}$ & \makecell[t{p{7cm}}]{ the radius indicating different interaction range, where $\omega=c,o,s,f$ represent the interaction, obstacle avoidance, safety, and observation radius, respectively. }\\
  $\mathcal{D}_{\omega}$ & \makecell[t{p{7cm}}]{ the dataset collected in different stages, where $\omega=c,~s,~e,~a$ represent the convergence, stabilization, excitation and attack processes, respectively.}\\
   $\mathcal{V}_{i}^{p}$ & the $r_i$-vertexed convex polygon covering $\mathcal{N}_{i}^{out}$\\  
 $\mathcal{N}_{i}^{\omega}$ & the in/out-neighbor set ($\omega=in/out$) of robot $i$ \\ 
 $\mathcal{V}_{\sss F}$ & the formation under partial observation \\  
 $\Omega$ & the rule set of formation interaction\\ 
   \bottomrule 
 \end{tabular} 
\end{table}
\subsection{Distance Constraints and Shape Specification}
The robots are assumed to be set in short distance interaction with their neighbors (by communication or sensing) and need to detect obstacles. 
The interaction radius $R_c$, the obstacle detection radius $R_o$, and the safe radius $R_s$ indicating danger range are supposed to satisfy 
\begin{equation}
R_c>R_o>R_s.
\end{equation}

Let $a_{ij}^0$ be the preset interaction weight of $(i,j)$ initially, and $z_i$ ($\bm{z_i}$) be the position state of robot $i$ in one (all) dimension(s). 
Based on the well-known disk model that considers the boundedness of $R_c$ \cite{bullo2009distributed}, $a_{ij}$ is assumed to satisfy
\begin{equation}\label{eq-12}
{a_{ij}} = a_{ij}^0 \cdot \text{sign}({R_c} - {\left\| {\bm{z_j} - \bm{z_i}} \right\|_2}), 
\end{equation}
where the indicative function $\text{sign}(x)\!=\!1$ if $x\!>\!0$, or $0$ otherwise. 
Accordingly, define the $R$-disk of arbitrary $\bm{z}$ as 
\begin{equation}
\mathcal{P}(\bm{z}, R)= \left\{ { \bm{z_{\omega}} : {{\left\| { \bm{z_{\omega}} - \bm{z} } \right\|}_2} < R} \right\}. 
\end{equation}


\subsection{Interaction Modeling of Formation Control} \label{s2-c}
The following modeling is given to describe how to form the preset shape through internal interaction, and deal with the environmental interferences through external interaction. 

\textit{Forming formation shape via internal interactions}. 
To describe the predefined geometric shape under formation control, a group of parameters $\{h_i\}(i\in\mathcal{V})$ is introduced, where $h_i$ is the desired relative deviation between robot $i$ (abbreviated to $r_i$ hereafter) and a common reference point. 
Note that once the formation shape is specified, the choice of the reference point will make no difference as $h_{ij}=h_j-h_i$ remains unchanged. 
To achieve this pattern, a commonly used first-order consensus-based formation control algorithm and its global form are given by \cite{olfati2007consensus}
\begin{equation}\label{eq-1}
{{\dot z}_i} \!\!=\!\!\!\sum\limits_{j \in \mathcal{N}_i^{in}} \!\!{{a_{ij}}(z_j\!-\!z_i\!-\!h_{ij})},~~\dot z(t) =- Lz(t)+Lh,
\end{equation}
where $h_{ij}=h_j-h_i$, $z = {\left[ { {{z_1}}, \cdots, {{z_{\sss N}}}} \right]^\mathsf{T}}$, and $h = {\left[ { {{h_1}}, \cdots, {{h_{\sss N}}}} \right]^\mathsf{T}}$. 
Generally, to dynamically guide the formation motion, one robot will be specified as the leader with an extra velocity input. 
For simplicity and without loss of generality, $r_{\sss N}$ is taken as the leader and reference node, and suppose that it runs in a constant velocity $c$. 
Denote $u_0=[0,\cdots,0, c]^\mathsf{T}$ and $u=Lh+u_0$. 
Then, the global dynamics in (\ref{eq-1}) is rewritten as 
\begin{equation}\label{eq-22}
\dot z(t)=- Lz(t)+u. 
\end{equation}
Accordingly, the solution of (\ref{eq-22}) is given by 
\begin{equation}\label{ll}
z(t) = {e^{ - Lt}}z(0) + \int_0^t {{e^{ - L(t - \tau )}}} ud\tau. 
\end{equation}

Note that (\ref{ll}) guarantees zero deviation for static formation and bounded tracking error for dynamic formation \cite{lewis2013cooperative}. 
To eliminate the error, solutions contain imposing a virtual leader to all robots (which is impractical when the scale of the MRN is large), or adopting a second-order controller which further utilizes the velocities of its neighbors. 
The global form by second-order controllers is the same as that by first-order ones, except that $z_i$ is augmented with the velocity state. 
Consequently, the internal interaction structure is related to both the positions and velocities of neighbors. 
To avoid fussy expressions and simplify the analysis, we adopt the first-order case in this paper. 
\begin{remark}
The second-order controller approximates the real leader speed, and can be seen as a nonlinear first-order model 
$\dot z(t)=- Lz(t)+Lh+u_s(t)$, where $u_s(t)$ is determined by the second-order controller design and satisfies $\mathop {\lim }\limits_{t\to \infty } u_s(t)= c\bm{1}$ \cite{lewis2013cooperative}. 
In Section \ref{simulation}, we will illustrate that our linear approximation method also applies to this nonlinear model if only using first-order states. 
\end{remark}

\textit{Obstacle avoidance via external interaction interface}. 
The obstacle-avoidance mechanism is the major interface for MRNs to interact with the physical environment. 
Let $z_{i*}$ be the desired goal of $r_i$,
$z_{ob}$ and $v_{ob}$ be the state and velocity of the obstacle, respectively.
The obstacle-avoidance behavior is mainly determined by the relative positions between a robot and the obstacle, and their velocities. 
Therefore, regardless of the detail design of different algorithms, we formulate the universal avoidance behavior as a general mechanism
\begin{equation}\label{local}
{{\dot z}_i} = g({z_{ob}}-{z_i},z_{i*}-{z_i},v_{ob},{v_i}). 
\end{equation}
Note that in multi-robot cases, $z_{i*}$ is usually time-varying and determined by ${\mathcal{N}_i^{in}}$, 
and numerous obstacle-avoidance mechanisms do not consider the mobility of obstacles, i.e., the last parameter in $g(\cdot)$ may not be used (e.g., in artificial potential method and genetic approach). 
Besides, the influence of $(z_{i*} - z_i )$ is negligible when $\left\| \bm{z_a}-\bm{z_i}\right \|_2 \to R_s$, and the effect of the obstacle is dominant. 
Typically, the boundedness of $g(\cdot)$ is described as follows
\begin{equation}
\left\{ {\begin{aligned}
&g( \cdot ) = 0,&&~\text{if}~\left\| \bm{z_{ob}}- \bm{z_i} \right \|_2> R_o,\\
&{0 \le |g( \cdot ) |\le b},&&~\text{if}~R_s\le\left\| \bm{z_{ob}}-\bm{z_i} \right \|_2\le R_o,\\
&{|g( \cdot )|=b},&&~\text{if}~\left\| \bm{z_{ob}}-\bm{z_i} \right \|_2<R_s.
\end{aligned}} \right.
\end{equation}

\textit{Formation maintenance}. 
To deal with possible faults and hazardous interference in the environment, many methods have been developed to support autonomy and self-organization for interaction maintenance, e.g., restoring the interaction topology of splitting robot groups \cite{liu2017self,li2019optimal}. 
In line with these considerations, we extract the main characteristics and formulate them as the following ruleset $\Omega=\{\Omega_R,\Omega_A\}$. 
\begin{itemize}
\item Neighbor-connection restoration rule $\Omega_R$. 
For $r_i$, if $a_{ij}(t)\ne a_{ij}^0$, $\forall j\in\mathcal{N}_i^{in}$ at time $t$, then $r_i$ will try to move back to its desired position $z_{i*}$ and reconnect with $r_j$, $\forall j\in\mathcal{N}_i^{in}$. 
The necessary condition of successful restoration is that $z_{i*}$ is available. 
\item Neighbor-replacement authentication rule $\Omega_A$. 
Given $r_a\notin \mathcal{V}$, $i\in \mathcal{V}$ and a time slot $t_l$, $j\in \mathcal{N}_i^{in}$ can be replaced by $r_a$ at $(t_0+t_l)$ if and only if $\forall t \!\in\! [{t_0},t_0+t_l)$, 
\begin{align}
a_{ij}(t)\ne a_{ij}^0,~~\bm{z_a}(t) \in \mathcal{P}(  \bm{z_i}(t), R_c),\\~~{\left\| {  \bm{z_a}(t) - \bm{z_{j*}}(t) } \right\|_2} \!<\! {\left\| {  \bm{z_j}(t) \!- \! \bm{z_{j*}}(t)} \right\|_2}. 
\end{align}
\end{itemize}

\begin{remark}
The reasonability and generality of the $\Omega$ lie in that it 
i) considers the prespecified neighborship configuration, instead of merely proximity among robots as in many existing works \cite{feng2016robust,zavlanos2009hybrid,kan2018distributed}; 
ii) works for situations where cyber authentication is not available, e.g., the interaction is purely based on sensors \cite{cheng2017event,aranda2015coordinate}. 
\end{remark}
The existence of the internal interaction structure, the external interaction mechanism, the organization ruleset $\Omega$ and their distance-constrained characteristic constitute all the weak knowledge of the attacker about the MRN. 
The knowledge is natural for an external attacker and provides no explicit information about system structure, model or parameters. 

\begin{figure}[t]
\centering
\setlength{\abovecaptionskip}{0.2cm}
\includegraphics[width=0.45\textwidth]{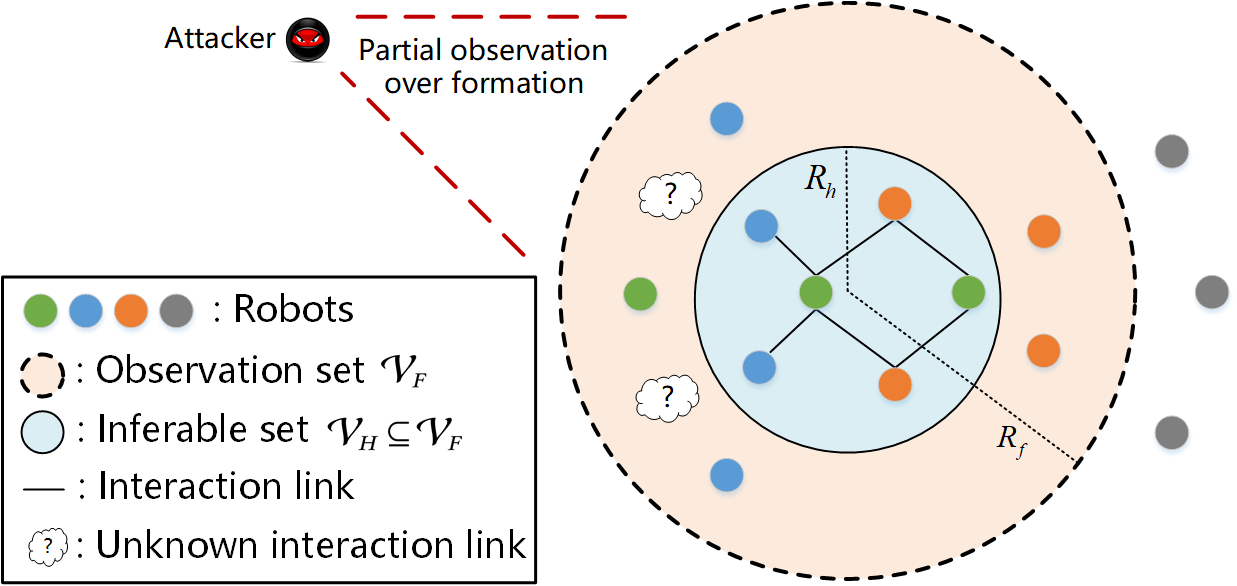}
\caption{Illustration of different observation ranges. 
The blue circle area enclosing $\mathcal{V}_{\sss H}$ is with radius $R_h$, and larger circle area enclosing $\mathcal{V}_{\sss F}$ is with radius $R_f$.}
\label{fig-range}
\vspace*{-8pt}
\end{figure}

\subsection{Partial Observation Ability of The Attacker}
Denote the sampling moment by $k$ and the discrete time form of (\ref{ll}) is given by 
\begin{equation} \label{basic-formation}
z^{k+1} = (I - \varepsilon_{\sss T} L)z^{k} + \varepsilon_{\sss T} u^{k} = Wz^{k} + \varepsilon_{\sss T} u^{k},
\end{equation}
where the sampling interval $\varepsilon_{\sss T}$ is small enough, and $W$ is the interaction structure matrix. 
Note that $W$ equivalently reflects the internal interaction structure and effect as $L$, since they are all linear transformation of the adjacent matrix $A$. 
Therefore, we directly seek to obtain $W$ for revealing the structure in the following sections.

Suppose the observation of the attacker is with \textit{i.i.d.} Gaussian noise $\xi_0^{k}\sim N(0,{\sigma ^2}I)$, and it is given by 
\begin{equation}\label{eq-3}
\tilde z^{k}=z^{k}+\xi_o^{k}=Wz^{k-1}+\varepsilon_{\sss T} u^{k}+\xi_o^{k}. 
\end{equation}
For every two consecutive observations, we obtain 
\begin{equation}\label{eq-4}
\begin{aligned}
\tilde z^{k}&=W(\tilde z^{k-1} -\xi_o^{k-1})+\varepsilon_{\sss T} u^{k}+\xi_o^{k}\\
&=W\tilde z^{k-1}+\varepsilon_{\sss T} u^{k}+\xi^{k},
\end{aligned}
\end{equation}
where $\xi=\xi_o- W\xi_o$. 
Note that (\ref{eq-4}) only illustrates the quantitative relationship between two consecutive measurements and does not represent the essential dynamic process.

Next, denote the partially observable part of $\mathcal{V}$ by $\mathcal{V}_{\sss F}$, and use the smallest enclosing circle covering $\mathcal{V}_{\sss F}$ with radius $R_f$. 
For convenience, we assume $R_f\ge2R_c$. 
Since a robot will only use information from their neighbors within its interaction range, 
for the robots near the observation boundary, their neighbors may not all be in $\mathcal{V}_{\sss F}$, making it hard to infer the interaction structure in $\mathcal{V}_{\sss F}$ with high accuracy. 
To deal with this issue, we narrow down the inferring range and use a concentric circle to cover the feasible subset $\mathcal{V}_{\sss H} \!\subseteq\! \mathcal{V}_{\sss F}$ with radius $R_h$, as shown in Fig.~\ref{fig-range}.

\begin{figure*}[t]
\centering
\setlength{\abovecaptionskip}{0.1cm}
\includegraphics[width=0.9\textwidth]{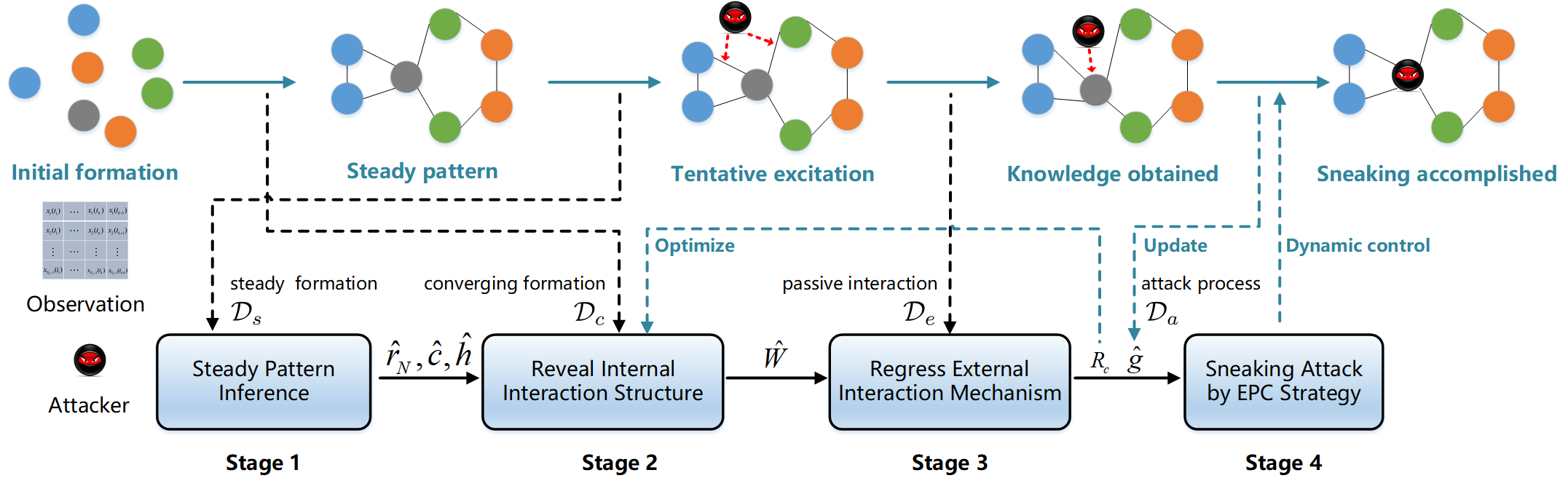}
\caption{The whole scheme of our proposed sneaking attack. 
First, at Stage 1 and 2, $r_a$ leverages $\mathcal{D}_c$ and $\mathcal{D}_s$ to infer the steady pattern parameters and $W$. 
At Stage 3, $r_a$ makes tentative excitation on $r_v$ and and regresses $g$ based on the observed response dataset $\mathcal{D}_e$. 
Meanwhile, the interaction radius $R_c$ is estimated and used as feedback to further optimize $W$. 
Finally, at Stage 4, $r_a$ begins its sneak with the support of the learned rules. And the response dataset $\mathcal{D}_s$ can also be used to expand the data for regressing $g$. 
}
\label{frame}
\end{figure*}

\subsection{Problem of Interest}\label{problemof}
An MRN $\mathcal{G}=(\mathcal{V},\mathcal{E})$ of $N$ robots is deployed to execute mission in physical environment, while running to a goal position $z_{g}$ with specified formation shape configuration $\{h_{ij}\}$.  
A malicious attack robot (denoted as $r_a$) who has limited observation over $\mathcal{V}_{\sss F}\subseteq \mathcal{V}$, aims to sneak into $\mathcal{V}$ by replacing a victim robot (denoted as $r_v$).

To achieve above purpose, $r_a$ needs to acquire sufficient interaction knowledge about $\mathcal{V}$. 
Note that what is most regular to $r_a$ is the steady formation shape, 
therefore, $r_a$ can first identity the steady pattern and then determine the convergence process, which reflects the internal interaction structure in $\mathcal{V}$. 
Since a robot in $\mathcal{V}$ has to react to the obstacles nearby, $r_a$ can actively make excitation on the robot to learn the external interaction rule. 
With the interaction rules obtained, $r_a$ begins the sneak attack efficiently. 
Accordingly, we divide the whole process into four stages: formation shape convergence (observe), steady formation maintenance (observe), tentatively excitation (explore), and attack launching (sneak). 
The observations on the four stages are constructed as four datasets: $\mathcal{D}_c$, $\mathcal{D}_s$, $\mathcal{D}_e$ and $\mathcal{D}_a$. 
Overall, to achieve the sneak attack, we need to solve the following four problems. 

\begin{itemize}
\item Stage 1: given $\mathcal{D}_s$, infer the steady pattern $\hat h$, $\hat c$.  
\item Stage 2: given $\mathcal{D}_c$, $\hat h$ and $\hat c$, find the mapping relation $\phi$ that embodies $W$ by solving 
\begin{equation}
\begin{aligned}
\mathop {\min }\limits_{\phi (\mathcal{D}_c) \mapsto W} {\| {W - \hat W} \|_{\sss \text{Frob}}}.
\end{aligned}
\end{equation}
\item Stage 3: given $\mathcal{D}_e$, find the mapping relation $\varphi$ that embodies $g$ by solving 
\begin{equation}
\begin{aligned}
\mathop {\min }\limits_{\varphi (\mathcal{D}_e) \mapsto g} {\| {g - \hat g} \|_2}. 
\end{aligned}
\end{equation}
\item Stage 4: given $\hat W$ and $\hat g$, design a control strategy $\bm{u}_{a,0:H}=\{u_a^0,u_a^1,\cdots,u_a^H\}$ such that
\begin{equation}
\mathcal{V}(t_{\sss H})=(\mathcal{V}(t_0)\backslash\{r_v\})\cup\{r_a\}.
\end{equation}
\end{itemize}
By solving the problems sequentially, $r_a$ is enabled to learn the interaction rules and leverage them to implement sneak attack. 
The whole framework with details is shown in Fig.~\ref{frame}. 

\begin{figure}[t]
\centering
\setlength{\abovecaptionskip}{0.1cm}
\includegraphics[width=0.45\textwidth]{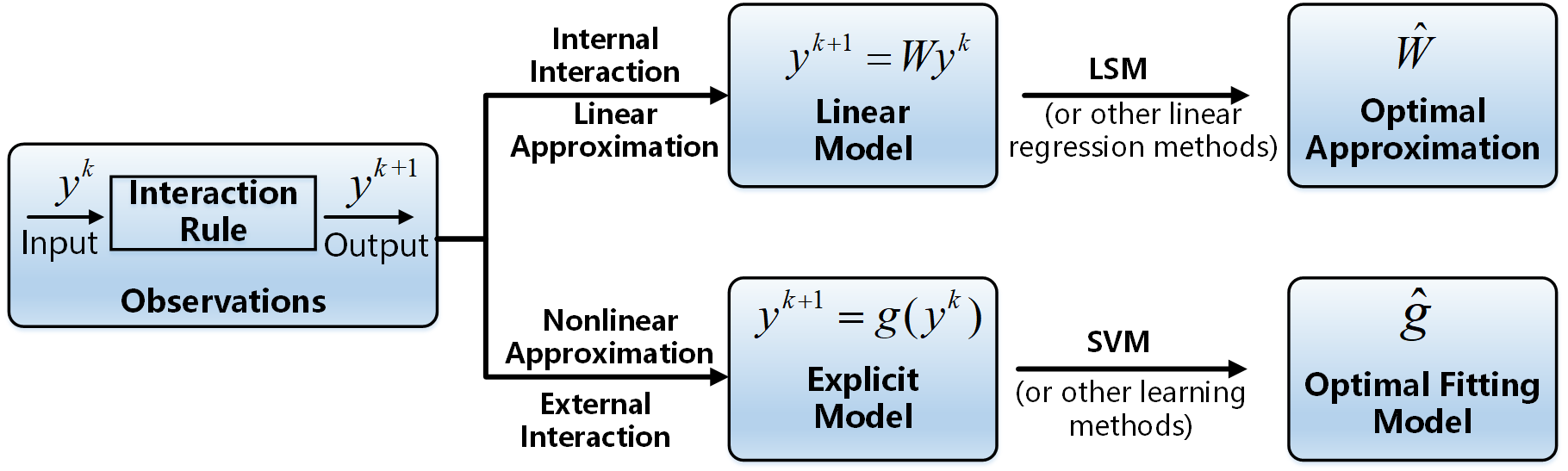}
\caption{Illustration of the approximation of the interaction rules. The formation state evolution is governed by the interaction rules, and the observed state at instant $k$ is a resulting output of the interaction rules taking the observed state at instant $k-1$ as input (possibly with preprocessing). }
\label{fig-app}
\vspace*{-5pt}
\end{figure}

\section{Revealing the Interaction Rules}\label{revealing}
In this section, we first analyze the feasibility of revealing the internal and external interaction rules of an MRN, and then highlight the methods addressing the closely related problems in Stage 1-3, respectively. 
Note that the interaction range $R_c$ is estimated by active excitation in Stage 3 and is used to approximate a more informative internal interaction structure than the one obtained in Stage 2. 
The obtained results are leveraged to constitute the cornerstone for the sneak attack. 

\subsection{Feasibility Analysis}
To reveal the interaction rules from the observations, there are two foremost issues that need to be addressed. 
First, under what condition the observations are meaningful and reflect the interaction rules. 
Second, how large the observations amount is sufficient to approximate the interaction rules with tolerable errors. 
If these issues are settled, the problem is then converted to parameter identification of a given system model or model fitting from a statistic perspective, and different methods are adopted to approximate the two kinds of interaction rules, as shown in Fig.~\ref{fig-app}. 
The details are given below.

For the internal interaction rule, what matters most is the structure among the robots. 
Therefore, the classical linear state space equation is sufficient to approximate the internal interaction rule, where the structure is explicitly represented as the state transition matrix. 
Under this model, the widely used least square method (LSM) provides an efficient way to obtain the optimal parameter estimation in the sense of minimizing the observation errors \cite{dobson2008introduction}. 
Note that we can only approximate $W$ locally due to the partial observation nature. 

For the external interaction rule, the obstacle-avoidance mechanism is fundamentally needed for a robot, and the avoidance behavior is inevitable if an obstacle is involved. 
Our major focus is to exploit how the robot will react to a close obstacle by $g$, not necessarily seeking the explicit analytical expression of $g$, especially considering its wide varieties. 
Therefore, we adopt generic learning-based methods to approximate the external interaction rule. 
Besides, due to the distance-constrained interaction, the interaction radius $R_c$ can be inferred through active excitation, and this result is utilized to expand the approximation range of a local $W$. 

\subsection{Steady Pattern Identification}
The steady pattern of the MRN reflects the formation shape and moving speed of the MRN. 

Assume that $r_a$ begins its observation as the MRN begins forming the formation shape from an unstable pattern, and $\mathcal{V}_{\sss H}$ is initially determined by a small $R_h$ (e.g., $R_h=0.3R_f$). 
When the formation pattern is reached, its state dynamics are illustrated by the following result.

\begin{theorem}\label{th01}
For the global dynamic model (\ref{ll}) with $u=Lh+[0\cdots0 \;c]^\mathsf{T}$, we have 
\begin{equation}
\mathop {\lim }\limits_{t \to \infty } \left\|z(t)-{c}t\! \cdot \!\bm{1} -s \right\|_2=0, 
\end{equation}
where the offset vector $s\!=\!{q_1}{p_1^\mathsf{T}}z(0)\!+\! (I-{q_1}{p_1^\mathsf{T}}) h\!+\!\sum\limits_{i = 2}^N { \frac{  cp_{\sss iN} }{{{\lambda _i}}}{q_i}}$. 
\end{theorem}

Theorem \ref{th01} illustrates that when the formation reaches the steady pattern, all robots are running at a common speed with fixed relative state deviations. 
This provides a theoretical foundation to identify the steady pattern parameters. 
Before that, let $\Delta z_i^k={z_i^k} \!-\! {z_i^{k-1}}$, $d_{ij}^k=\|{z_j^k}-{z_i^k}\|_2$, and we define the $l$-period window as $[k,k+l]$. 
The second-order state difference accumulation of $r_i$ during $l$-period is given by
\begin{equation}
{  \Delta S_{i}^{k_0:k_0+l} } \!=\! \sum\limits_{k = k_0+1}^{k_0+l-1} {{\| \Delta z_i^{k+1} - \Delta z_i^{k} \|}_2} .
\end{equation}
Then, based on the observations, the steady velocity is obtained by solving 
\begin{equation} \label{c0}
\hat c(k_0,l) = \mathop {\arg \min }\limits_c \sum\limits_{k = {k_0}}^{{k_0} + l} {{{\left\| {\tilde z_{\sss F}^k} - (c{\varepsilon _T}k + {b_0})\bm{1} \right\|}_2}},
\end{equation}
where $b_0$ is an auxiliary constant. 
Note that (\ref{c0}) can be easily solved by generic LSM. 
Next, define the real time when the steady pattern is reached as $k_s$, and we present the following theorem to show how to identify the formation leader and the performance of estimated $\hat c$. 
\begin{corollary}\label{thooo}
Given  $\mathcal{V}_{\sss F}$ and $k_s$, if the leader $r_{\sss N}\in \mathcal{V}_{\sss F}$, then it is identified by 
\begin{equation}
\hat r_{\sss N}= {\arg \mathop {\min }\limits_{i} \left\{ \Delta  S_{i}^{2:k_s}: \Delta S_{i}^{2:k_s}<\epsilon, i\in\mathcal{V}_{\sss F} \right\}},  \label{ww1}
\end{equation}
where $\epsilon$ is an arbitrarily small positive constant. 
$\forall k_0>k_s$, we have 
$\mathop {\lim }\limits_{l \to \infty } \hat c({k_0},l) = c. $
\end{corollary}

Corollary \ref{thooo} is straightforward due to the uniqueness of $r_{\sss N}$ and the steady pattern characteristic. Thus the proof is omitted here. 
Note that smaller $\Delta S_{i}^{k_0:k_0+l} $ means less velocity variation in $r_i$, and it illustrates how to identify the leader and estimate the stable velocity. 
Based on it, given preset $l$ and $\epsilon$, the following criterion is designed to estimate $k_s$, i.e.,
\begin{equation}\label{time}
{\hat k_s} \!=\! \inf \left\{ {{k_0} \!:\left({  \sum\nolimits_{i \in {\mathcal{V}_{\sss F}}}  \Delta \tilde S_{i}^{k_0:k_0+l} } \right)\!\le\! \epsilon } \right\}.
\end{equation}
Once $r_a$ obtains $\hat k_s$, the datasets $\mathcal{D}_c$ are $\mathcal{D}_s$ are constructed by 
\begin{equation}\label{data}
\mathcal{D}_c=\{ \tilde z_{\sss F}^k:k \le \hat k_s \},~\mathcal{D}_s=\{ \tilde z_{\sss F}^k:k > \hat k_s \},
\end{equation}
where $\tilde z_{\sss F}^k=[\tilde z_{i_1}^k,\tilde z_{i_2}^k,\cdots, \tilde z_{i_{|{\sss{ \mathcal{V}_{\sss F} }}|}}^k]^\mathsf{T}$. 
Meanwhile, the formation shape parameters $\{h_i\}$ are calculated by 
\begin{equation}
\hat h = \hat s - {{\hat s}_{j}} \bm{1}, 
\end{equation}
where $\hat s = {  \sum\nolimits_{k = \hat k_s+1}^{\hat k_s+l} {(\tilde z_{\sss F}^k - \hat c{\varepsilon _{\sss T}} k\!\cdot\! \bm{1})} }/{l}$.

\subsection{Internal Interaction Structure Approximation}
After identifying the steady pattern, we then try to extract the interaction structure among $\mathcal{V}_{\sss F}$.

Denote the unobserved robot set as $\mathcal{V}_{\sss F'}=\mathcal{V}\backslash\mathcal{V}_{\sss F}$, and define the indicative leader vector 
\begin{equation}
\mathbb{I}_{\sss F}^{[i]}=\left\{ {\begin{aligned}
&1, &&\text{if}~\exists i\in \mathcal{V}_{\sss F}, i=\hat r_{\sss N},\\
&0, &&\text{otherwise}. 
\end{aligned}} \right.
\end{equation}
Then, the global dynamics (\ref{eq-4}) is divided into 
\begin{equation}
\left[ {\begin{aligned}
{z}_{\sss F}^{k+1}\\
{z}_{\sss F'}^{k+1}
\end{aligned}} \right] \!=\! \left[ {\begin{aligned}
W_{\sss FF}~W_{\sss FF'}\\
W_{\sss F'F}~W_{\sss F'F'}
\end{aligned}} \right]\left[ {\begin{aligned}
{z}_{\sss F}^{k}\\
{z}_{\sss F'}^{k}
\end{aligned}} \right] \!+ \!{\varepsilon _{\sss T}}\left[ {\begin{aligned}
u_{\sss F}^{k}\\
u_{\sss F'}^{k}
\end{aligned}} \right],
\end{equation}
where $u_{\sss F}=h_{\sss F}+ c \mathbb{I}_{\sss F}$ and $u_{\sss F'}=h_{\sss F'}+c \mathbb{I}_{\sss F'}$. 
Accordingly, the observations of $r_a$ over $\mathcal{V}_{\sss F}$ is given by
\begin{equation} \label{eq-6}
{{\tilde z}_{\sss F}}^{k+1} = {W_{\sss FF}}{{\tilde z}_{\sss F}}^{k} + {W_{\sss FF'}}{{\tilde z}_{\sss F'}}^{k} + \varepsilon _{\sss T}\hat u_{\sss F}^{k}+\xi_{\sss F}^{k}.
\end{equation}
Based on (\ref{eq-6}), there are two points that need to be noticed: 
\begin{itemize}
\item For $r_a$, it can only observe ${z}_{\sss F}$, unaware of ${W_{\sss FF'}}$ and ${z}_{\sss F'}$;
\item The evolution of ${z}_{\sss F}$ is a coupled consequence of ${z}_{\sss F}$ itself and the unobservable ${ z}_{\sss F'}$.
\end{itemize}
Therefore, it is quite difficult to directly solve ${W_{\sss FF}}$ by (\ref{eq-6}).

To avoid the above issue, we narrow down the inference range from $\mathcal{V}_{\sss F}$ to a smaller $\mathcal{V}_{\sss H}$. 
Let $\mathcal{V}_{\sss H'}=\mathcal{V}_{\sss F}\backslash{\mathcal{V}_{\sss H}}$, 
$W_{\sss HF}=[W_{\sss HH}~W_{\sss HH'}]$, $y_{\sss H}^{k}=\tilde z_{\sss H}^{k} -\hat h_{\sss H} - {\varepsilon _{\sss T}}\hat c\mathbb{I}_{\sss H} $ and 
$y_{\sss F}^{k} = {[(\tilde z_{\sss H}^{k} -\hat h_{\sss H} )^\mathsf{T},(\tilde z_{\sss H'}^{k})^\mathsf{T}]^\mathsf{T}}$. 
We obtain the following result of approximating the structure. 
\begin{theorem}\label{th-topo}
Given $\mathcal{D}_c$ and ${\hat k_s}$, if $\mathcal{D}_c$ is linearly modeled by $r_a$, then in the sense of expectation, the observations satisfy 
\begin{equation}\label{tt}
\bm{E}(y_{\sss H}^{k+1}) = {W_{\sss HF}}\bm{E}(y_{\sss F}^{k}).
\end{equation}
If $|\mathcal{V}_{\sss F}| \! + \! 1\! \le |\mathcal{D}_s|$,
then the optimal estimation of ${\phi (\mathcal{D}_c)\mapsto W}$ in sense of least square is
\begin{equation}\label{solution1}
\phi (\mathcal{D}_c):~\hat W_{\sss HF}=\left(  ({Y_{\sss F}} {Y_{\sss F}^\mathsf{T}})^{^{-1}} {Y_{\sss F}}  {Y_{\sss H}^\mathsf{T} }   \right)^\mathsf{T}, 
\end{equation}
where ${Y_{\sss H}} = [y_{\sss H}^{2},y_{\sss H}^{3},\cdots,y_{\sss H}^{l}]$ and 
${Y_{\sss F}} = [y_{\sss F}^{1},y_{\sss F}^{2},\cdots,y_{\sss F}^{l-1}]$.
\end{theorem}

Theorem \ref{th-topo} gives the least square solution of $W_{\sss HF}$ under the linear approximation model. 
Although the number of feasible observations is limited in practice, Theorem \ref{th-topo} nevertheless can be used as the basis for approximating $W_{\sss HF}$, especially to obtain a robust connection structure even the real model is nonlinear and with different noise variance, which will be verified in Section \ref{simulation}. 

\subsection{External Interaction Rule Approximation}\label{estimate_external}
Next, we present the tentative excitation based method to illustrate how to infer the interaction radius $R_c$ and construct dataset $\mathcal{D}_e$ for learning the external interaction rule $g$.

\begin{definition}\label{direct}
(\textbf{Direct controllability}) 
A node is directly controllable if one can control it to reach any given state $z_c^*$ in finite steps by direct external excitations.
\end{definition}
\begin{lemma}\label{attack1}
If $g$ and $z_{i*}$ is known, $\forall i \in \mathcal{V}_{\sss F}$, $r_i$ is directly controllable by $r_a$. 
\end{lemma}

Lemma \ref{attack1} shows that $r_i$ can be steered to any state by utilizing the characteristic of $g$. 
Given the input configuration of $g$, the avoidance behavior is unique. 
Therefore, as in \cite{li2019learning}, we make $r_a$ actively excite on $r_i$ and observe its external interaction response to learn $g$. 

First, based on the analysis in last section, for $i\in \mathcal{V}_{\sss F}$, its desired position of $z_{i*}$ is estimated by 
\begin{equation}\label{predict}
\hat z_{i*}^{k+1}\!=\! 
\left\{ 
\begin{aligned}
&\hat c\cdot k+\hat s^{[i]},~~~~~\text{if}~\mathcal{N}_i^{in}~\text{is~in~steady~pattern}; \\
&\sum\limits_{j \in\mathcal{V}_{\sss F} } {{{\hat a}_{ij}}({\tilde z_j^k} - {\tilde z_i^k}-  h_{\sss F}^{[j]}+ h_{\sss F}^{[i]} )},~\text{otherwise},
\end{aligned}\right.
\end{equation} 
where ${\hat a_{ij}} =\hat w_{ij}/{\varepsilon _{\sss T}}~(i\neq j)$. 
The next implicit issue is the obstacle detection radius $R_o$, which can be directly inferred by approaching $r_i$ from a distant position.

Then, the excitation strategy is designed considering two possible situations. 
If there $\exists i\in\mathcal{V}_{\sss H}$, $|\mathcal{N}_i^{in}|>1$, we randomly select a $j\in\mathcal{N}_i^{in}$ as the direct excitation target. 
Note that the excitation is adopted when $\mathcal{N}_j^{in}$ is in steady pattern, thus $z_{j*}$ is available. 
The excitation input is designed such that $u_e$ is orthogonal to the ideal moving direction of $r_j$, $\overrightarrow{dir}(\bm{z_{j*}})$, i.e., 
\begin{equation}
\!\{ {\bm{u_e}}\!:{\bm{u_e}} \!=\! \lambda \overrightarrow{dir}^{\perp}(\bm{z_{j*}}),\|\bm{u_e}\|_2\!\le\! b, {\left\| {{\bm{z_a}}({\bm{u_e}})\! - \!{\bm{z_{j*}}}} \right\|_2} \!<\! {\hat R_o}\},\!\! 
\end{equation}
where $\lambda>0$ and ${\perp}$ means the orthogonal vector. 
Since $r_i$ is an out-neighbor of $r_j$, when within in the interaction range, its state is predicted by 
\begin{equation}
{{\hat z}_i}^{k+1} = {  \hat W_{\sss HF}^{[i,:]} }{{\tilde z}_{\sss F}}^{k} +\varepsilon _{\sss T}\hat u_i^{k}. 
\end{equation}
Once the continuous excitation of $r_a$ makes $r_i$ lose connections with any one in $\mathcal{N}_i^{in}$, 
the deviation $\| \bm{\hat{z}}_{i}^{k+1}- \bm{\tilde{z}}_{i}^{k+1}\|_2$ will be significant large, which inspires us to design the following criterion to estimate $R_c$, given by
\begin{equation}\label{aa}
\left\{ 
\begin{aligned}
    k_2&=\min\{k:\left\|  \bm{\hat z_i}^{k+1}- \bm{\tilde z_i}^{k+1} \right\|_2>\beta\}, \\
    \hat R_c&=\max\{\| \bm{\tilde z_i}^{k_2}-  \bm{\tilde z_j}^{k_2}\|_2:~j\in\mathcal{N}_i^{in}\}, 
\end{aligned}\right.
\end{equation}
where $\beta>0$ is a given threshold. 
Technically, a larger $\beta$ will guarantee a more conservative $\hat R_c$. 

If there $\forall i\in\mathcal{V}_{\sss H}$, $|\mathcal{N}_i^{in}|\le1$, we randomly select a $r_i$ with $|\mathcal{N}_i^{in}|=1$ as the direct excitation target. 
In this case, we introduce a time window $k_l$ to design an intermittent excitation strategy. 
At each iteration $k$ satisfying $[k\!\!\mod\! k_l=0]$, $ \bm{u_e}(k)$ is randomly selected from 
\begin{equation}
\!\{ \bm{u_e}\!:\bm{{z_a}({u_e}}) \!=\! \alpha \bm{z_{i*}} \!+\! (1 \!-\! \alpha )\bm{z_i},{\left\| { \bm{{z_a}({u_e}}) \!-\! \bm{z_{i*}} } \right\|_2} \!<\! {\hat R_o}\},\!
\end{equation}
and $r_a$ stays still at other iterations. 
Likewise, the same judging criterion (\ref{aa}) is used to estimate $R_c$ at iteration $k$ satisfying $[k\mod k_l=k_l-1]$. 
Once a $\hat R_c$ is estimated, $r_a$ does reverse excitation to steer the robot to form the original steady pattern, and observes more obstacle-avoidance behaviors.

Finally, with $\hat R_c$ determined, $\mathcal{V}_{\sss H}$ is re-specified by setting $R_h=R_f-\hat R_c$, and a more informative $\hat W_{\sss HF}$ is further approximated by (\ref{solution1}). 
During the whole process of this stage, the avoidance behavior corresponding to every active excitation is recorded as the following input-output pair
\begin{equation}\label{input1}
\left\{ 
\begin{aligned}
&Q_{in}^{k}\!=\![\tilde z_v^{k}-\tilde z_a^{k}, \tilde z_{v*}^{k}-\tilde z_v^{k}, \Delta \tilde z_v^{k}/\varepsilon_{\sss T},\Delta \tilde z_a^{k}/\varepsilon_{\sss T}],\\
&Q_{out}^{k}=\Delta \tilde z_v^{k+1}.
\end{aligned}\right.
\end{equation}
Further, construct the dataset $\mathcal{D}_e = \left\{ {\mathop  \cup \limits \{Q_{in}^{k},Q_{out}^{k}\} } \right\}$, and $g$ can be learned using many mature learning methods by solving 
\begin{equation}\label{ques}
\hat g = \mathop {\arg \min }\limits_{g: Q_{in}\mapsto Q_{out}} \sum\nolimits_{k = 1}^{ |\mathcal{D}_e | } {{{\left\| Q_{out}^{k}-g(Q_{in}^{k}) \right\|}_2}}.
\end{equation}
Specifically, we adopt SVR method, which has good performance on nonlinear approximation and strong generalization ability when the amount of data is not vast \cite{cristianini2000introduction}. 

At the end of this stage, $\hat R_c$, $\hat W_{\sss HF}$ and $\hat g$ are all obtained, providing support to design efficient sneak strategies.

\section{Sneak Attack Design}\label{attack}
In this section, we first analyze the attack feasibility by introducing the definitions of 1-hop convex formation and the indirect controllability of a robot. 
Then, the sufficient conditions for the attack are established. 
Finally, we propose the ECR strategy to dynamically replace $r_v$ by $r_a$, completing sneak.

\subsection{Feasibility Analysis}
In this part, we give the theoretical guarantees for the existence of attack positions and the controllability of robots. 

Note that the success of the attack lies in two parts: i) there exists a feasible attack position and ii) $r_v$ is controllable such that the attack condition is satisfied. 
Accordingly, we introduce the following definitions. 
\begin{definition} \label{polygon}
(\textbf{1-hop convex pattern}) 
Given $r_i$, if $\exists \mathcal{V}_{i}^{p}\subseteq\{ i \cup \mathcal{N}_i^{out} \}$ such that i) the nodes in $\mathcal{V}_{i}^{p}$ constitutes a convex polygon; 
ii) the polygon covers $\{ i \cup \mathcal{N}_i^{out} \}$; 
iii) $i$ is one vertex of the polygon, then $\mathcal{V}_{i}^{p}$ is the 1-hop convex pattern of $r_i$. 
\end{definition} 
\begin{definition}
(\textbf{Indirect controllability}) 
A node is indirectly controllable if one can control another node, and a let a chain reaction make the tagged node reach any $z_c^*$ in finite steps. 
\end{definition}
\begin{theorem}\label{property}
Given $r_i$, if $\mathcal{V}_{i}^{p}$ exists, then
\begin{enumerate}[i)]
\item $\mathcal{V}_{i}^{p}$ is unique. 
\item $\mathcal{Z}_i^{f}={ \mathop \bigcap \limits_{\mathclap{j \in \mathcal{N}_i^{out} }} {\mathcal{P}}( \bm{z_j} , d_{ij})}\neq\emptyset$, and ${\left\| { \bm{z} - \bm{z_j} } \right\|_2} < {\left\| { \bm{z_i} - \bm{z_j} } \right\|_2}$, $\forall \bm{z}\in \mathcal{Z}_i^{f}$, $ j \in \mathcal{N}_i^{out}$. 
\end{enumerate}
\end{theorem}

Theorem \ref{property} shows the nice distance properties brought by $\mathcal{V}_{i}^{p}$, which helps to the sneak strategy design. 
Besides, in its proof in Appendix, we also provide a method to find the positions that satisfy the property. 
Next, we illustrate under what conditions the robots are indirectly controllable. 

Suppose that $i\in\mathcal{V}$ is under the excitation of $r_a$, and it can be seen as another leader. 
Then, $z_i$ is directly determined by $u_e=g(\cdot)$. 
In this situation, denote the new adjacent matrix as $A^e$ where 
$a^e_{ij}=0,\forall j\in\mathcal{N}_i^{in}$, and other elements are the same as in $A$. 
Accordingly, its Laplacian matrix is $L^e=\text{diag}\{A^e \bm{1}\}-A^e$, and $p^e_1\!=\![p^e_{\sss 11},\!\cdots\!,p^e_{\sss 1N}]^\mathsf{T}$ is the corresponding left eigenvector for $\lambda_1$ of $L^e$.
\begin{lemma}\label{lh44}
Given desired state $z_c^*$ and initial state $z_i^0$, $r_i$ is indirectly controllable by $r_j$ iff the following conditions hold
\begin{equation}
\left\{ 
\begin{aligned}
&u_e u_c>0,&\text{if}~(z_c^*-z_i^0)u_c>0, \\
&|p^e_{\sss 1j}{u_e}| >|p^e_{\sss 1N}{u_c}|,&\text{if}~(z_c^*-z_i^0)u_c<0. 
\end{aligned}\right.
\end{equation}
\end{lemma}

Lemma \ref{lh44} points out the necessary and sufficient condition for indirect controllability from the perspective of global connectivity. 
However, under partial observation, the eigenvectors of the interaction matrix are hard to obtain, making it impossible to verify the conditions. 
To overcome this issue, we leverage the idea of Lemma \ref{lh44}, and obtain the local conditions that guarantee the direct controllability. 
\begin{theorem}\label{th44}
Given desired state $z_c^*$ and initial state $z_i^0$, $r_i$ is indirectly controllable by $r_j$ when $u_e,u_c$ satisfy 
\begin{equation}\label{str}
\left\{ 
\begin{aligned}
&u_e u_c>0,&\text{if}~(z_c^*-z_i^0)u_c>0, \\
&|a_{ij}{u_e}| >|\bar{a}_{ij}{u_c}|,&\text{if}~(z_c^*-z_i^0)u_c<0,
\end{aligned}\right.
\end{equation}
where $\bar{a}_{ij}=\!\!\!\!\! \sum\limits_{j' \in \{ \mathcal{N}_i^{in}\backslash{j} \}}{\!\!\!\!\!  a_{ij'}}$. 
\end{theorem}
Without knowing the global interaction structure in $\mathcal{V}$, Theorem \ref{th44} provides a sufficient criterion to find the robots satisfying the indirect controllability by considering the local interaction with their in-neighbors. 
This condition will be used in the following strategy design.

\subsection{Sneak Strategy: Evaluate-Cut-Restore}
To achieve maximum attack rewards, a feasible $r_v$ needs to be selected based on its significance evaluation in the formation first, and then $r_a$ is controlled to replace $r_v$. 
Besides, the strategy design should be flexible for different local geometric patterns of $r_v$. 
To meet the neighbor-replacement rule $\Omega_A$, the key point is to ensure the existence of $\mathcal{V}_{v}^{p}$. 
Briefly, the ideas of our proposed sneak strategy are summarized as follows. 
\begin{itemize}
\item \textbf{Evaluate} the significance of the robots based on their potential indirect controllability and the state-propagation effects. 
Select the most valuable one as $r_v$. 
\item \textbf{Cut} the connections between $r_v$ and $\mathcal{N}_v^{in}$. 
If the 1-hop convex pattern $\mathcal{V}_{v}^{p}$ does not exist, $r_a$ needs to make $\mathcal{V}_{v}^{p}$ form first. 
\item \textbf{Restore} the original steady pattern, by making $r_a$ be regarded as the in-neighbor of $\mathcal{N}_v^{out}$ based on ruleset $\Omega$ and replacing $r_v$ by $r_a$.
\end{itemize}

During the sneak stage, the estimated state of $r_v$ under attack and the state of $r_a$ are respectively given by  
\begin{equation}\label{state1}
\hat z_v^{k + 1} (u_a^k)= \tilde z_v^k + \hat g({u_a^k}), 
\end{equation}
\begin{equation}\label{state2}
z_a^{k+1}=z_a^k+u_a^k,
\end{equation}
where $\hat g({u_a^k})$ is simplified for $\hat g({z_a^k}-{\tilde z_i^k},{z_{i*}^k}-{\tilde z_i^k},{v_a^k}-{v_i^k})$. 
The details of ECR strategy are given below. 

\textbf{Evaluate phase}. 
With $\hat W_{\sss HF}$ obtained, we need to find a victim robot to achieve the maximum rewards. 
Note that for a robot in the MRN, a larger out-degree means it has the potential capability to control more robots while a smaller in-degree indicates it is less affected by others. 
From this perspective, the evaluation criteria is designed as 
\begin{subequations}\label{node-value0}
\begin{align}
\mathop {\max }\limits_{i}  ~~~~~~&( |\mathcal{N}_i^{out}|+ {\| {W_{\sss\! HF}^{[:,i]}} \|_1} - |\mathcal{N}_i^{in}|-{\| {W_{\sss HF}^{[i,:]}}\|_1})\\
\text{s.t.}~~~~~~&i \in \mathcal{V}_{\sss H}, ~\mathcal{N}_i^{out}|\ge1, |\mathcal{N}_i^{in}|\le \alpha_1,
\end{align}
\end{subequations}
where ${W_{\sss HF}^{[:,i]}}$, ${W_{\sss HF}^{[i,:]}}$ represent the $i$-th column and row of $W_{\sss HF}$, respectively, and $\alpha_1$ a preset possible integer. 
The solution of (\ref{node-value0}) is selected as $r_v$. 
Due to the small scale of the observed formation, (\ref{node-value0}) can be easily solved by a common traverse search method with algorithm complexity of $O(|\mathcal {H}| |\mathcal {F}|)$.

\textbf{Cut phase}. 
In this core phase, $r_a$ aims to make the connections between $r_v$ and its in-neighbors break, i.e., $a_{vj}(t)\ne a_{vj}^0$. 
If $\mathcal{V}_{i}^{p}$ exists and $r_v$ is readily to be excited directly (i.e., not many other robots are distributed around), the following strategy is adopted to break the connections,
\begin{subequations}\label{eq-pro}
\begin{align}
\mathop{\max }\limits_{\bm{u_a}^k}~&{\alpha_2}{\| {\bm{\hat z_v}^{k + 1}(\bm{u_a}^k) \!-\! \bm{\hat z_{v*}}^{k + 1}}\|_2} + {\alpha _3}\!\sum\limits_{j \in \mathcal{N}_v^{in}} \!\!\!{{{\| {\bm{\hat z_j}^{k + 1} \!-\! \bm{\hat z_v}^{k + 1} \!-\! {\bm{\tilde h_{jv}} }} \|}_2}} ,\label{eq-pro-a}\\
{\rm{s.t.}}~& (\ref{predict}),~(\ref{state1})~\text{for}~r_v,~\text{and}~(\ref{state2}),\\ 
&\|\bm{u_a}^k \|_2 < b.  
\end{align}
\end{subequations}
where $\alpha_2,\alpha_3>0$ are the weight parameters to balance the offsets of $r_v$ with its ideal position and in-neighbors. 
An efficient way to solve this cut-edge problem is to adopt a greedy heuristic by sampling from the feasible solution space. 
The attack iteration under (\ref{eq-pro}) is terminated by the same criterion (\ref{aa}) for estimating $R_c$ in Section \ref{estimate_external}. 

If $\mathcal{V}_{v}^{p}$ does not exist or the $r_v$ can not be attacked readily, we utilize the indirect controllability to attack an in-neighbor of $r_v$ first, such that the formation pattern $\mathcal{V}_{i}^{p}$ is reached and $r_v$ is easier to approach. 
Let $r_{l_{1}}, r_{l_{2}}$ be the two robots in $\mathcal{N}_i^{out}$ that are closest to $r_j$, 
and $\overrightarrow{\bm{z}_{l_1} \bm{z}_{l_2} }$ be the vector from $\bm{z}_{l_1}$ to $\bm{z}_{l_2}$. 
Then, the following strategy is adopted
\begin{align}\label{chain}
\bm{u_a}^k = \mathop {\arg }\limits_{\bm{u_a}} \max \{ &{{\| {\bm{\hat z_j}^{k + 1}(\bm{u_a}) - \bm{\hat z_{j*}}^{k + 1}} \|}_2}: \bm{u_a}\in \mathcal{U}_{\sss F}, \nonumber \\
&j\in\mathcal{N}_i^{in} ~\text{and}~ r_j~\text{satisfies}~(\ref{str})  \},
\end{align}
where $\mathcal{U}_{\sss F}\!=\!\{\bm{u_a}: { {\left\| {\bm{u_a}} \right\|}_2} < b,\bm{u_a} = \lambda \overrightarrow{\bm{z}_{l_1} \bm{z}_{l_2} }^{\perp} ,\forall \lambda>0 \} $. 
Iteratively, $r_a$ updates the control inputs until $\mathcal{V}_{i}^{p}$ is formed, then the control mode is turned to (\ref{eq-pro}).

\textbf{Restore phase}. 
After the $\mathcal{V}_{v}^{p}$ is formed and the connections between $r_v$ and $\mathcal{N}_v^{in}$ are cut, $r_a$ only needs to be in a position that is close to arbitrary $j\in\mathcal{N}_v^{in}$. 
Utilizing the properties in Theorem \ref{property}, $r_a$ is controlled by 
\begin{align}\label{chain2}
\bm{u_a}^k = \mathop {\arg }\limits_{\bm{u_a}} \max \left\{ {{\left\| {\bm{\hat z_v}^{k + 1}(\bm{u_a}) - \bm{\hat z_{v*}}^{k + 1}} \right\|}_2}: \bm{z_a}^{k+1}\in \mathcal{Z}_v^{f}  \right\}.
\end{align}
By continuously implementing the procedure, $r_a$ takes the ideal position that belongs to $r_v$, and will be regarded as the real $r_v$ by $r_v$'s out-neighbors at certain time according to rule $\Omega_A$. 
After that, the original $r_v$ becomes an outlier and $r_a$ is controlled to move towards $\hat z_{v*}$ and $\bm{z_a} \in \mathcal{P}(\bm{z_j}, R_c)$. 
Finally, $r_a$ is recognized as a member and obtains partial control over $\mathcal{V}$. 
The sneak attack is completed. 

During this attack-implementing stage, $r_a$ also collects dataset $\mathcal{D}_a=\{Q_{in}^{k},Q_{out}^{k}\}$ and uses it to update $\hat g$ online. 
After sneaking into the MRN, $r_a$ is able to implement further attacks, for example, infer the interaction structure in $\mathcal{V}_{\sss H'}$, or intentionally split the MRN by misleading $\mathcal{N}_v^{out}$ and the out-neighbors of them. 
These interesting problems are not the focus of this paper and are omitted here. 

To sum up, the whole procedures of the sneak attack consisting of four stages are summarized in Algorithm \ref{algo-all}. 

\begin{algorithm}[t]
 \small
    \caption{Sneak Attack against MRNs}
    \label{algo-all}
    \begin{algorithmic}[1]
    \REQUIRE{Partial observation range $R_f$, moving ability bound $b$.}\;
    \ENSURE{Interaction rules $W_{\sss HF}$, $g$.}\;
    \STATE Observe the dynamic process of $\mathcal{V}_{\sss F}$ converging to the steady formation geometric pattern.\;
    \STATE Judge when the steady pattern is reached by (\ref{time}).\;
    \STATE Construct datasets $\mathcal{D}_c$, $\mathcal{D}_s$, and estimate the parameters of steady pattern by (\ref{data}).\;
    \STATE Obtain an initial approximation of $W_{\sss HF}$ based on $\mathcal{D}_c$ by (\ref{solution1}).\;
    \STATE Leveraging direct controllability, make tentative excitation on $r_j$ continuously to infer $R_o$, $R_c$ by (\ref{aa}).\;
    \STATE Expand the approximation range of $W_{\sss HF}$ utilizing $\hat R_c$.\;
    \STATE Construct dataset $\mathcal{D}_e= \left\{ {\mathop  \cup \limits \{Q_{in}^{k},Q_{out}^{k}\} } \right\}$ during the excitation process, and learn the external interaction mechanism $g$ by (\ref{ques}).\;
    \STATE Leveraging indirect controllability, adopt ECR (Evaluate-Cut-Restore) attack strategy to replace a selected $r_v$ by $r_a$.\;
 	\STATE Construct dataset $\mathcal{D}_a= \left\{ {\mathop  \cup \limits \{Q_{in}^{k},Q_{out}^{k}\} } \right\}$ during the sneak process, and update $g$ online  by (\ref{ques}).\;
 	\STATE Obtain partial control over $\mathcal{V}_{\sss F}$ and support further attacks.\;
    \end{algorithmic}
\end{algorithm}

\section{Simulation}\label{simulation}
In this section, we conduct extensive simulations to demonstrate the feasibility of revealing the interaction rules from external observations and excitation, and also validate the theoretical analysis and the proposed sneak attack strategy.

\subsection{Simulation Setting}
\begin{figure}[t]
  \setlength{\abovecaptionskip}{0.1cm}
    \subfigure[]{ 
    \label{sae1} 
    \includegraphics[width=0.23\textwidth]{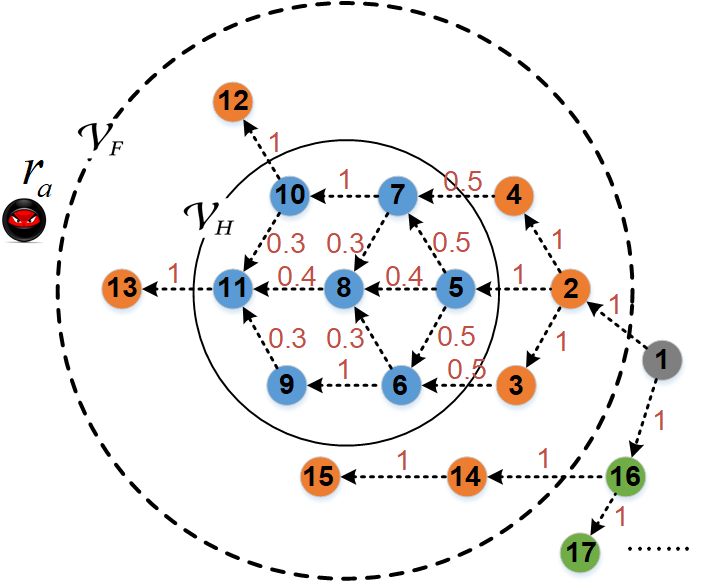} 
  }
  \hspace{-5mm}
  \subfigure[]{ 
    \label{sae2} 
    \includegraphics[width=0.23\textwidth]{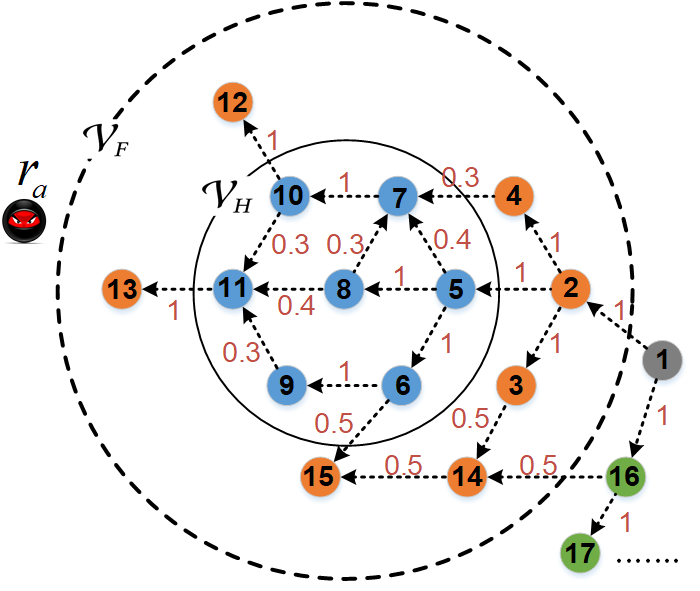} 
  } 
  \caption{An MRN of 17 robots with two different interaction structure. The formation shape is set the same, while the intersection structure is different. 
  The detailed interaction weights are denoted in red font. } 
  \label{se_example}
  \vspace*{-13pt}
\end{figure}

\textbf{MRN specification}. 
An MRN consisting of 17 robots is considered, with two kinds of interaction structure and common preset formation shape, as shown in Fig.~\ref{se_example}. 
Specifically, robot 1 is set as the leader, and the moving speed in stable stage is set as $0.2m/s$. 
The maximum speed of $r_a$ is $1m/s$, and the radii are setting as $R_c=7m$, $R_o=2m$ and $R_s=0.5m$. 

\begin{figure*}[ht]
\setlength{\abovecaptionskip}{0.1cm}
  \centering 
    \subfigure[]{ 
    \label{se0} 
    \includegraphics[width=0.3\textwidth]{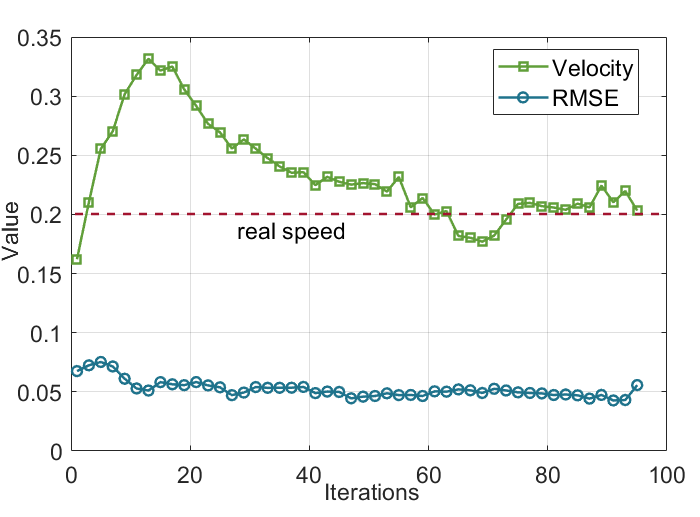} 
    }
    \subfigure[]{ 
    \label{se1} 
    \includegraphics[width=0.3\textwidth]{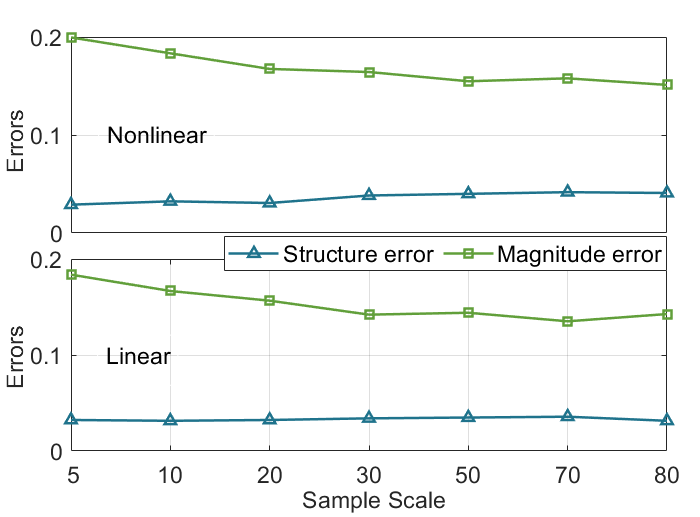} 
  } 
  \subfigure[]{ 
    \label{se2} 
    \includegraphics[width=0.3\textwidth]{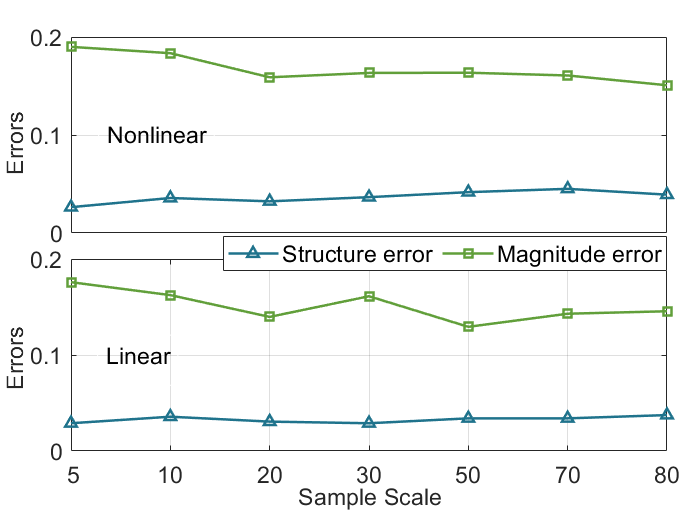} 
   } 
  \caption{ The approximation results based on calculated $\hat k_s$. (a). Formation velocity estimation. (b) and (c) are the approximation results with increasing data amount of $\mathcal{D}_c $ structure case 1 and 2, respectively.} 
  \label{error_sample}
  \vspace*{-8pt}
\end{figure*}

\textbf{Comparisons under different models and parameters}. 
For the interaction structure $\hat W_{\sss HF}$, we present the approximation results from three perspectives. 
First, we adopt the liner model $\dot z(t) =- Lz(t)+Lh+u_0$ and the nonlinear model $\dot z(t) =- Lz(t)+Lh+u_s(t)$, respectively. 
Second, we consider the influence of different sample amounts and different observation noise variance. 
Additionally, $\hat W_{\sss HF}$ can be further optimized by considering the feedback of $\hat R_c$, i.e., 
\begin{subequations} \label{further}
\begin{align}
\mathop {\min }\limits_{{W_{\sss HF}}} ~& \| Y_{\sss H} - {W_{\sss HF}}Y_{\sss F}  \|_{\sss \text{Frob}}  \\
\text{s.t.}~~&W_{\sss HF}^{[i,j]}=0,\text{if}~\|\tilde z_i-\tilde  z_j\|_2>\hat R_c, i\in\mathcal{V}_{\sss H},j\in\mathcal{V}_{\sss F}.
\end{align}
\end{subequations}
As a typical constrained linear least square problem, (\ref{further}) can be solved by many mature optimization techniques, e.g., Newton interior-point method \cite{boyd2004convex}. 
For the external interaction mechanism $\hat g$, we present the approximation results under different sample amounts and different noise variance.

\textbf{Metric of evaluation}. 
To evaluate $\hat W_{\sss HF}$, we use the following two indexes 
\begin{align}
{\varepsilon _1} &=({{{\| {\text{sign}({\hat W}_{\sss HF}) - \text{sign}(W_{\sss HF}}) \|}_0}})/  {( |\mathcal{V}_{\sss H}| |\mathcal{V}_{\sss F}| )}, \\
{\varepsilon _2} &= ({\| {{\hat W}_{\sss HF} - W_{\sss HF}} \|_{\text{Frob}} }) / {\| W_{\sss HF}\|_{\text{Frob}}},
\end{align}
where ${\varepsilon _1}$ evaluates the structure correctness in terms of whether two robots have an interaction connection, and ${\varepsilon _2}$ represent the magnitude correctness in terms of how larger is the connection weight between two robots. 
To evaluate $\hat g$, we adopt mean directional accuracy (MDA), root mean square error (RMSE) and mean absolute error (MAE), which are respectively calculated by  
\begin{align}
{\rm{MDA}} &= \frac{1}{m}\sum\limits_{i = 1}^m {  { \text{sign}  }({y_i} - {{y'}_i}} ),\\
{\rm{RMSE}} &= \sqrt {\frac{1}{m}\sum\limits_{i = 1}^m {{{({y_i} - {{y'}_i})}^2}} } ,\\
 {\rm{MAE}} &= \sum\limits_{i = 1}^m {\frac{{\left| {{y_i} - {{y'}_i}} \right|}}{{{m}}}},
\end{align}
where $y_i$, $y'_i$ are the actual and predicted values, $m$ is the amount of samples. 

\subsection{Results and Analysis}
We mainly illustrate the results from the perspectives of the four stages described in Section \ref{problemof}. 

\textbf{Stage 1}. 
Recall that the time instant $\hat k_s$ is computed by (\ref{time}) to identify when the MRN has reached the steady pattern. 
Using different threshold $\epsilon$ to compute $\hat k_s$ will result in different sample number of $\mathcal{D}_c $. 
As shown in Fig.~\ref{se0}, when the MRN reaches steady state, the velocity estimation remains stable. 
Fig.~\ref{se1}-\ref{se2} illustrate the sample number mainly affects the accuracy of $\varepsilon_2$, and when the amount becomes larger, both $\varepsilon_1$ and $\varepsilon_2$ become stable. 
This is because that the latter observations in $\mathcal{D}_c $ are observed from the stage that is close to the steady process, only providing redundant information to approximate ${W}_{\sss HF}$. 
In fact, this demonstrates that an accurate calculation of $\hat k_s$ is not important under both linear and nonlinear models, especially for the structure error. 
Nevertheless, more data are still helpful to obtain ${W}_{\sss HF}$ with smaller magnitude errors.

\begin{figure}[t]
  \centering 
  \setlength{\abovecaptionskip}{0.1cm}
    \subfigure[]{ 
    \label{error_w_a} 
    \includegraphics[width=0.24\textwidth]{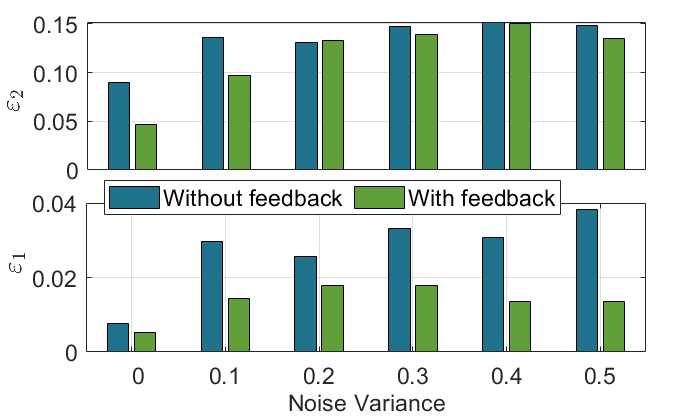} 
  } 
  \hspace{-6mm}
  \subfigure[]{ 
    \label{error_w_b} 
    \includegraphics[width=0.24\textwidth]{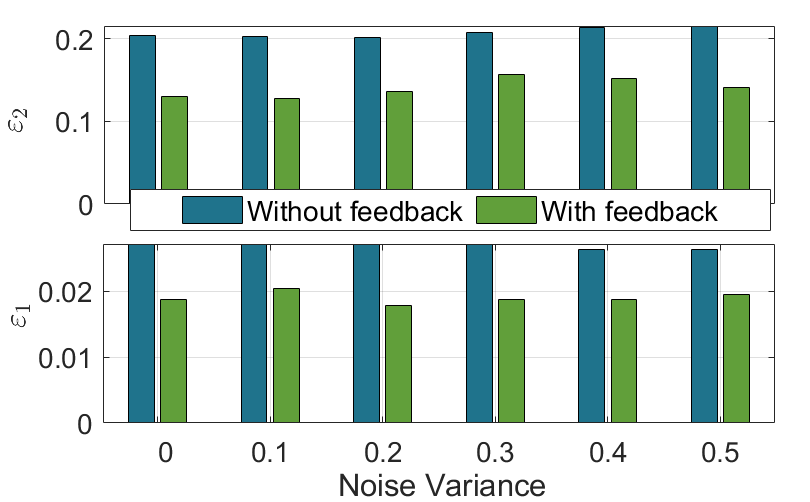} 
  } 
  \caption{The approximation result comparison of ${\hat W}_{\sss HF}$ with and without the feedback of $\hat R_c$. (a) under linear formation control. (b) under nonlinear formation control.} 
  \label{error_w}
  \vspace*{-10pt}
\end{figure}

\textbf{Stage 2}. 
Hereafter, we present the results based on the interaction structure given in Fig.~\ref{sae1}. 
Fig.~\ref{error_w} shows the approximation errors of the internal interaction structure under different variances of observation noises. 
The variance of observation noise varies from 0 to 0.5. 
Fig.~\ref{error_w_a} presents the approximation evaluation of the MRN under linear formation law. 
As it shows, the structure error $\varepsilon_1$ is smaller than $\varepsilon_2$, and when noises are involved, the approximation error is generally stable under different noise variances. 
Specifically, when the interaction radius feedback is used to further optimize ${W}_{\sss HF}$, i.e., solving (\ref{further}), the approximation errors are reduced significantly. 
The details in Fig.~\ref{error_w_b} is likewise, and under the same conditions other than the nonlinear formation control, the approximation errors are basically larger than that under linear formation control, which is not hard to understand. 
However, we observe that even though the approximation errors in the nonlinear case is higher, the difference is not significant. 
The errors remain small and are sufficient to get a feasible ${W}_{\sss HF}$ to support the following attack, especially in terms of the connection structure. 
These results verify that the accuracy of the proposed linear approximation for ${W}_{\sss HF}$. 
\begin{figure*}[t]
\centering
\setlength{\abovecaptionskip}{0.05cm}
\includegraphics[width=0.9\textwidth]{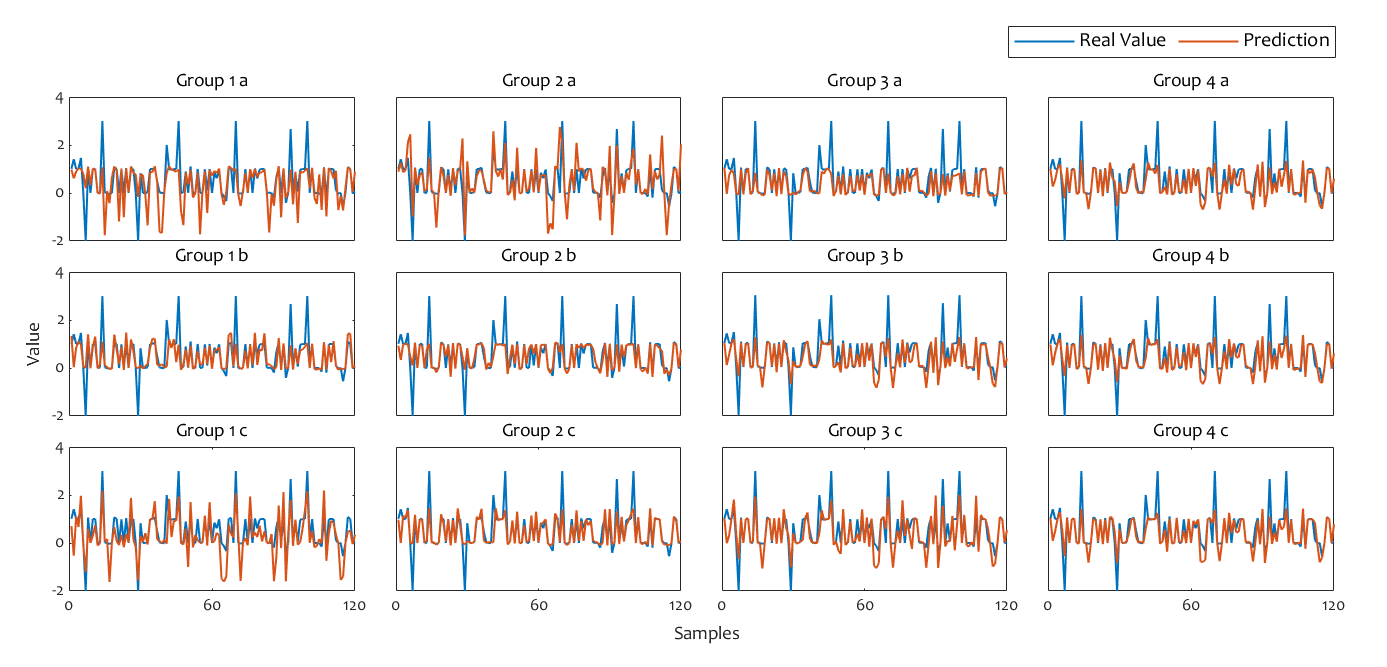}
\caption{Obstacle-avoidance mechanism regression results based on SVR method. 
Under different numbers of the samples, 3 groups of experiments are conducted for each sample number. }
\label{regression_noise}
\vspace*{-10pt}
\end{figure*}

\textbf{Stage 3}. 
To verify the approximation performance of the obstacle-avoidance mechanism (here we use classical artificial potential method \cite{khatib1986real}), SVM with a Gaussian kernel function is used for regression, and the parameters are optimized by 10-fold grid search cross-validation. 
We conduct 4 groups of regression with 3 repeated procedures based on $\mathcal{D}_e$ collected from the tentative excitation process. The $\sigma=0.1$ here. 
We randomly select 25, 50, 100, 200 samples to train the SVR model in each group, respectively. 
A fixed set of 120 samples are used for testing. 
The testing results are shown in Fig.~\ref{regression_noise}, 
with the statistic results given in Table ~\ref{regression_table}. 
By comparing the results of the test samples in each group, it is obvious that the fitting performance is improved with more training data. 
Meanwhile, the index does not vary much with more samples, which is illustrated by comparing the results of repeated tests. 
The MDA result shows a rising trend from 25-sample-training to 200-sample-training, which indicates that more training samples lead to high accuracy. 
Nevertheless, we observe that even though a larger number of samples bring more accurate results, the improvement of the accuracy is not significant. 
Even with a small number of samples such as the 25-samples, the method is still feasible, especially in terms of MDA (above 90\%). 
It is worth mentioning that RMSE and MAE fluctuate due to observation noises. 
Specifically, RMSE is no larger than 0.6, corresponding to the famous $3\sigma$-principle \cite{pukelsheim1994three}. 
Overall, these show the effectiveness of the proposed method to learn the external interaction rule.

\begin{table}[]
\setlength{\abovecaptionskip}{0.3cm}
\caption{Statistic obstacle-avoidance mechanism regression results based on SVR method.  }
	\begin{tabular}{cccclccc}
		\toprule[1pt]
		& \multicolumn{3}{c}{25 samples}  &  & \multicolumn{3}{c}{50 samples}  \\ \cline{2-4} \cline{6-8} 
		Index & MDA       & RMSE     & MAE      &  & MDA       & RMSE     & MAE      \\ \cline{2-8} 
		Training         & 0.880        & 0.253    & 0.154    &  & 0.913      & 0.217    & 0.113    \\
		Testing          & 0.933      & 0.601    & 0.404    &  & 0.933      & 0.581    & 0.300    \\ \hline
		& \multicolumn{3}{c}{100 samples} &  & \multicolumn{3}{c}{200 samples} \\ \cline{2-4} \cline{6-8} 
		Index & MDA       & RMSE     & MAE      &  & MDA       & RMSE     & MAE      \\ \cline{2-8} 
		Training         & 0.910        & 0.333    & 0.146    &  & 0.923      & 0.426    & 0.206    \\
		Testing          & 0.956      & 0.541    & 0.291    &  & 0.967      & 0.496    & 0.264    \\ 
		\bottomrule[1pt]
	\end{tabular}
		\label{regression_table}
		\vspace*{-10pt}
\end{table}

\begin{figure}[t]
  \centering 
    \subfigure[]{ 
    \label{stage_a} 
    \includegraphics[width=0.4\textwidth]{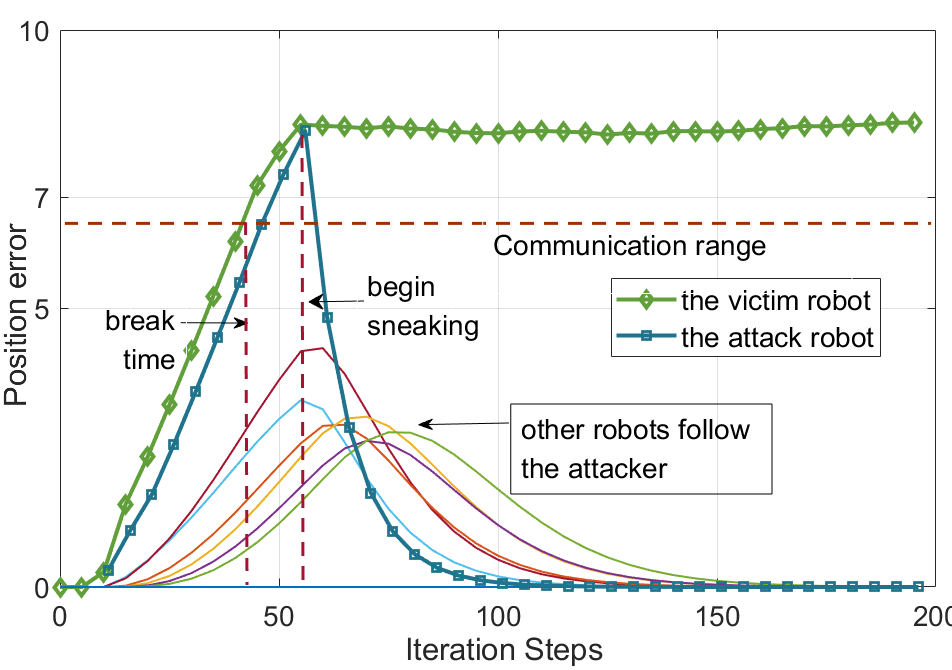} 
  } 
  \subfigure[]{ 
    \label{stage_b} 
    \includegraphics[width=0.4\textwidth]{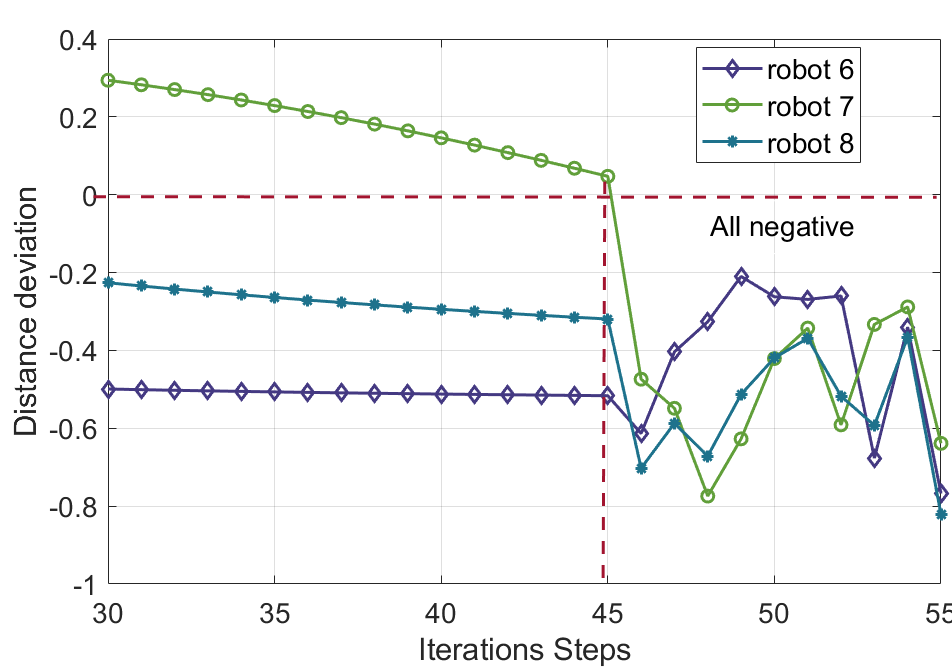} 
  } 
  \caption{The illustration of the sneak process. (a). The position errors between the real and the desired positions of the robots, and $r_a$ takes the $z_5^*$ as its desired position. (b) The distance deviations  $({\left\| {z_a - \tilde z_j} \right\|_2} - {\left\| {{\tilde z_i} - \tilde z_j} \right\|_2}) $, here $i=5$, $j=6,7,8$.  } 
  \label{attacker_w}
  \vspace{-15pt}
\end{figure}

\textbf{Stage 4}. 
First, at the Evaluate phase, robot 5 is selected as $r_v$ by the evaluation criteria (\ref{node-value0}), which has only one in-neighbor and three out-neighbors, and its corresponding 1 hop convex pattern $\mathcal{V}_{5}^{p}$ exists. 
Then, during the Cut phase, $r_a$ utilizes the learned $\hat g$ and makes consequent excitations on $r_v$ by solving (\ref{eq-pro}), such that $r_v$ is gradually pulled out of the interaction range of its in-neighbors, shown as the phase before the break time in Fig.~\ref{stage_a}. 
After the connections of $r_v$ with its in-neighbors break, $r_a$ in the Restore phase is controlled by solving $(\ref{chain2})$ dynamically to move within $\mathcal{Z}_i^{f}={ \mathop \bigcap \limits_{\mathclap{j \in \mathcal{N}_i^{out} }} {\mathcal{P}}(z_j, d_{ij})}$ ($i=5$ here), such that ${\left\| {z_a - z_j} \right\|_2} < {\left\| {{z_i} - z_j} \right\|_2}$, $\forall j \in \mathcal{N}_i^{out}$ (i.e., $j=6,7,8$ here). 
This phase is illustrated as the process between the break time and sneaking start time in Fig.~\ref{stage_a}, with detailed presentation of distance deviations $({\left\| {z_a - \tilde z_j} \right\|_2} - {\left\| {{\tilde z_i} - \tilde z_j} \right\|_2}) $ in Fig.~\ref{stage_b}, where the deviations are all negative. 
Based on the neighbor-replacement authentication rule $\Omega_A$, the robots $\mathcal{N}_5^{out}$ will take $r_a$ as the new neighbor replace robot 5. 
Finally, $r_a$ here just moves towards $z_{5*}$ and its out-neighbors begin following it, and the formation will reach steady state again. 
Note that the position fluctuation of $\mathcal{N}_5^{out}$ before beginning sneaking in Fig.~\ref{stage_a} also verifies the indirect controllability in Theorem \ref{th44}. 
Likewise, this can be used to intentionally split the MRN formation shape, and the results are omitted here due to space limits.

\section{Conclusion}\label{conclusion}
In this paper, we investigated the security of MRNs under formation control by designing a sneak attack. 
We demonstrated that the interaction rules in MRNs are learnable even the attacker is only with partial observation over the MRN, without any strong prior knowledge like the system dynamics model or structure. 
First, we proposed an excitation-based method to approximate the internal interaction structure within the MRN and the obstacle-avoidance mechanism with the environment. 
Then, we designed the ECR strategies, which make the attacker replace a victim robot and gain optimal local control over the MRN. 
Finally, comprehensive theoretical analysis and numerical results illustrated the effectiveness of the proposed attack. 
This work reveals the possibility of learning the interaction rules of an MRN from physical observations and excitations, and shows how to utilize them to design an intelligent attack, even though the interaction rules are model agnostic to the attacker. 

In a broader sense, our work broadens the horizons of the security of CPSs by exploiting the system vulnerabilities from physical observations and interactions. 
Furthermore, we emphasize that the system interaction nature can be leveraged by malicious attackers to cause disastrous consequences on the system. 
Concerning the CPS security from this perspective, a lot of issues remain to be further addressed, including i) how to design efficient detection methods to identify the potential threats; 
ii) how to reliably secure the interaction protocols that only leak confusing states on the external observer; 
and iii) how to construct robust and dynamic interaction defensive strategies with the physical world.

\bibliographystyle{IEEEtran}

\section*{Appendix}\label{appen}
\subsection{Proof of Theorem \ref{th01}}
\begin{proof}
Note ${e^{ - Lt}} = \Gamma{e^{ - \Lambda t}}{\Gamma^{ - 1}}$, and substituting it into (\ref{ll}), the global state $z(t)$ is decomposed as 
\begin{align}\label{eq-th11}
z(t) =& \sum\limits_{i = 1}^N {{q_i}{p_i^\mathsf{T}}{e^{ - {\lambda _i}t}}} z(0) + \int_0^t {\sum\limits_{i = 1}^N {{q_i}{p_i^\mathsf{T}}{e^{ - {\lambda _i}(t - \tau )}}} } d\tau  \cdot u \nonumber \\
=&{q_1}{p_1^\mathsf{T}}z(0) + \sum\limits_{i = 2}^N {{q_i}{p_i^\mathsf{T}}{e^{ - {\lambda _i}t}}} z(0)\nonumber \\
 &+\left[ {{q_1}{p_1^\mathsf{T}}t + \sum\limits_{i = 2}^N {\frac{1}{{{\lambda _i}}}{q_i}{p_i^\mathsf{T}}(1 - {e^{ - {\lambda _i}t}})} } \right] \cdot u.
\end{align}
Then, define an auxiliary variable ${z_{ss}} = {q_1}{p_1^\mathsf{T}}u \cdot t +{q_1}{p_1^\mathsf{T}}z(0) +  \sum\limits_{i = 2}^N {\frac{1}{{{\lambda _i}}}{q_i}{p_i^\mathsf{T}}u}$ and one infers that 
\begin{align} \label{f22}
\mathop {\lim }\limits_{t \to \infty } \left\|z(t)-{z_{ss}} \right\|_2=0. 
\end{align}

Next, since $p_i^\mathsf{T}q_j\!=\!0~(i\!\ne \!j)$ and $\lambda_1=0$, substituting $L=(\sum\limits_{i = 1}^N {{q_i}p_i^T{\lambda _i}} )$ into ${q_1}{p_1^\mathsf{T}}L$ and $\sum\limits_{i = 2}^N {\lambda _i}{q_i}{p_i^\mathsf{T}}L$, it yields that
\begin{equation} \label{f33}
\left\{ {\begin{aligned}
&{q_1}{p_1^\mathsf{T}}(\sum\limits_{i = 1}^N {{q_i}{p_i^\mathsf{T}}{\lambda _i}} )=0, \\
&\sum\limits_{i = 2}^N {{q_i}{p_i^\mathsf{T}}(\sum\limits_{i = 1}^N {{q_i}{p_i^\mathsf{T}}} )}= \sum\limits_{i= 2}^N {{q_i}{p_i^\mathsf{T}}}= (I \!-\! {q_1}{p_1^\mathsf{T}}).
\end{aligned}} \right.
\end{equation}
Note $u = Lh + {u_0} = (\sum\limits_{i = 1}^N {{q_i}p_i^T{\lambda _i}} )h + {u_0}$, substituting (\ref{f33}) into  ${z_{ss}}$, we further obtain
\begin{equation}
{z_{ss}} = {q_1}p_1^T{u_0} \cdot t + {q_1}p_1^Tz(0) + (I - {q_1}p_1^T) + \sum\limits_{i = 2}^N {\frac{1}{{{\lambda _i}}}{q_i}p_i^T{u_0}}.
\label{eqadd1}
\end{equation}

Since $r_{\sss N}$ is a unique leader and does not use information from others, the last row of matrix $A$ and $L$ are zeros. 
Then, the following equations hold
\begin{equation}\label{f44}
\left\{ {\begin{aligned}
&{p_1^\mathsf{T}L =[p_{\sss 11},\cdots,p_{\sss 1N}]^\mathsf{T}L =[0,\cdots,0]},\\
&{{p_1^\mathsf{T}}{q_1} =[p_{\sss 11},\cdots,p_{\sss 1N}]^\mathsf{T}\bm{1} = 1}.
\end{aligned}  } \right.
\end{equation}
By solving (\ref{f44}), we obtain
\begin{equation}\label{f55}
\left\{ {\begin{aligned}
&{{p_{\sss 1N}} = 1},\\
&{{p_{\sss 1k}} = 0,k = 1,2, \cdots ,N\! -\! 1}.
\end{aligned}} \right.
\end{equation}
Substituting (\ref{f55}) into (\ref{eqadd1}) and it follows that
\begin{equation}
{z_{ss}} = {q_1}{p_1^\mathsf{T}}z(0)+ (I-{q_1}{p_1^\mathsf{T}}) h+\sum\limits_{i = 2}^N { \frac{  cp_{\sss iN} }{{{\lambda _i}}}{q_i}}  + {c}t\! \cdot \!\bm{1} = s + {c}t\! \cdot \!\bm{1}
\end{equation}
Finally, it is derived that $\mathop {\lim }\limits_{t \to \infty } \left\|z(t)-{c}t\! \cdot \!\bm{1} -s \right\|_2=0$.
\end{proof}


\subsection{Proof of Theorem \ref{th-topo}}
\begin{proof}
Here we adopt the limit analysis method. 
First, if $R_f=R_c$ (i.e., the circle center locates on a robot $i$), 
then ${\mathcal{N}_i^{in}  \subseteq \mathcal{V}_{\sss F}}$, which yields
\begin{equation}\label{qq}
{{\tilde z}_i}^{k+1} = {  W_{\sss FF}^{[i,:]} }{{\tilde z}_{\sss F}}^{k} +\varepsilon _{\sss T}\hat u_i^{k}+ \xi_{i}^{k}, 
\end{equation}
where $W_{\sss FF}^{[i,:]}$ represents the corresponding row for $z_i$. 
By moving terms, it follows that 
\begin{equation}\label{eq-single}
 {  W_{\sss FF}^{[i,:]} }{{\tilde z}_{\sss F}}^{k}={{\tilde z}_i}^{k+1} -\varepsilon _{\sss T}\hat u_i^{k} - \xi_{i}^{k}={y_i^{k+1}} - \xi_{i}^{k}.
\end{equation}
Supposing the observations are noise-free and no other prior information is available, then $|\mathcal{V}_{\sss F}|$ groups of observation equations are sufficient to solve $W_{\sss HF}^{[i,:]}$. 
This case shows that robot $r_i$ is only influenced by its in-neighbors that are within its communication range. 

Next, if $R_f>R_c$, then $\exists i\in \mathcal{V}_{\sss H} \subseteq \mathcal{V}_{\sss F} $, $\{i\cup\mathcal{N}_i^{in}\} \subseteq \mathcal{V}_{\sss F}$. 
Therefore, (\ref{qq}) is generalized as 
\begin{align}\label{eq-c}
{\tilde z_{\sss H}}^{k+1} &= {W_{\sss HH}}{\tilde z_{\sss H}}^{k} + W_{\sss HH'}\tilde z_{\sss H'}^{k}+\varepsilon _{\sss T}\hat u_{\sss H}^{k}+\xi_{\sss H}^{k} \nonumber \\
&=W_{\sss HF}\tilde z_{\sss F}^{k}+\varepsilon _{\sss T} \hat u_{\sss H}^{k}+\xi_{\sss H}^{k}.
\end{align}
Substitute $\hat u_{\sss H}=\hat h_{\sss H}+\hat c \mathbb{I}_{\sss H}$ into (\ref{eq-c}) and it follows that 
\begin{align}
{{\tilde z}_{\sss H}}^{k+1}&=W_{\sss HF}\tilde z_{\sss F}^{k}+{\varepsilon _{\sss T}}\{ \hat c \mathbb{I}_{\sss H} + \frac{{I - {W_{\sss HH}}}} {\varepsilon _{\sss T}}\hat h_{\sss H}\} \nonumber \\
&= {W_{\sss HH}}({{\tilde z}_{\sss H}}^{k} - \hat h_{\sss H}) + {W_{\sss HH'}}{{\tilde z}_{\sss H'}}^{k} + {\varepsilon _{\sss T}} \hat c \mathbb{I}_{\sss H} + \hat h_{\sss H} +\xi_{\sss H}^{k}.
\end{align}
By the definition of $y_{\sss H}$ and $y_{\sss F}$, we obtain
\begin{align}
\label{pf-noise}
y_{\sss H}^{k+1}  &={{\tilde z}_{\sss H}}^{k+1}- {\varepsilon _{\sss T}} \hat c \mathbb{I}_{\sss H} - \hat h_{\sss H} \nonumber \\
&= {W_{\sss HH}}({{\tilde z}_{\sss H}}^{k} - \hat h_{\sss H}) + {W_{\sss HH'}}{{\tilde z}_{\sss H'}}^{k} +\xi_{\sss H}^{k}\nonumber \\
& = {W_{\sss HF}}y_{\sss F}^{k}+\xi_{\sss H}^{k}. 
\end{align}
Since $\bm{E}(\xi_{\sss H}^{k})=0$, we have $\bm{E}(y_{\sss H}^{k+1}) = {W_{\sss HF}}\bm{E}(y_{\sss F}^{k})$.

Finally, to obtain the least square solution of ${W_{\sss HF}}$ from the observations, we ignore the noise term in (\ref{pf-noise}). 
Note that a single group $y_{\sss H}^{k+1}={W_{\sss HF}}y_{\sss F}^{k}$ is based on two consecutive observations over $\mathcal{V}_{\sss F}$, and contains $|\mathcal{V}_{\sss H}|$ groups of equations about $W_{\sss HF}$, which consist of $(|\mathcal{V}_{\sss H}|\!\cdot\!|\mathcal{V}_{\sss F}|)$ parameters. 
For $i\in\mathcal{V}_{\sss H}$, the transpose of $y_{\sss i}^{k+1}={W_{\sss HF}^{[i,:]}}y_{\sss F}^{k}$ is given by 
\begin{equation}
y_{i}^{k+1}=(y_{\sss F}^{k})^\mathsf{T} (W_{\sss HF}^{[i,:]})^\mathsf{T}. 
\end{equation}
Stack the states of continuous moments into one and it yields 
\begin{equation}
\left[ {\begin{aligned}
&y_{i}^{2}\\
&~\vdots\\ 
&y_{i}^{l}
\end{aligned}} \right] \!=\! \left[ {\begin{aligned}
&(y_{\sss F}^{1})^\mathsf{T}\\
&~~\vdots\\ 
&(y_{\sss F}^{l-1})^\mathsf{T}
\end{aligned}} \right](W_{\sss HF}^{[i,:]})^\mathsf{T}.
\end{equation}
Then, one easily infers the least square solution of the vector $(W_{\sss HF}^{[i,:]})^\mathsf{T}$. 
Integrating all robots in $\mathcal{V}_{\sss H}$, it follows that 
\begin{equation}
\label{pf-topo}
Y_{\sss H}^\mathsf{T}=
\left[ {\begin{aligned}
&(y_{\sss H}^{2})^\mathsf{T}\\
&~\vdots\\ 
&(y_{\sss H}^{l})^\mathsf{T}
\end{aligned}} \right] \!=\! \left[ {\begin{aligned}
&(y_{\sss F}^{1})^\mathsf{T}\\
&~~~\vdots\\ 
&(y_{\sss F}^{l-1})^\mathsf{T}
\end{aligned}} \right]W_{\sss HF}^\mathsf{T}=Y_{\sss F}^\mathsf{T}W_{\sss HF}^\mathsf{T}. 
\end{equation}
To avoid a under-determined solution of $W_{\sss HF}$, at least $(|\mathcal{V}_{\sss H}|+1)$ consecutive observations over $\mathcal{V}_{\sss F}$ are needed to solve (\ref{pf-topo}). 
The final estimation is given by $\hat W_{\sss HF}=\left(  ({Y_{\sss F}} {Y_{\sss F}^\mathsf{T}})^{^{-1}} {Y_{\sss F}}  {Y_{\sss H}^\mathsf{T} }   \right)^\mathsf{T}$. 
The proof is completed. 
\end{proof}

\subsection{Proof of Lemma \ref{attack1}}
\begin{proof}
Due to $r_a \notin\mathcal{V}_{\sss F}$, when $\left\|z_a-z_i\right \|_2<R_o$, $r_a$ is regarded as an external obstacle by $r_i$. 
Note that in cases where $\left\|z_a-z_i\right \|_2\le R_s$, the influence of $(z_{i*}-z_i)$ is neglected since $r_i$ aims to move as far from $r_a$ as it could. 
By Definition \ref{direct}, given the goal state $z_c^*$, the proof turns to find a groups of inputs $\bm{u_a}=\{u_a^k,k=1,\cdots k_{\sss H}\}$ to satisfy 
\begin{equation}
\begin{aligned}
z_c^* \!-\! z_i^0 &= \sum\limits_{k = 1}^{{k_H}} {g\left( {z_a^k(u_a^k) \!-\! z_i^k,z_{i*}^k \!-\! z_i^k,v_a^k, v_i^k} \right)}.
\end{aligned}
\end{equation}
With $g$ known and $(z_{i*}-z_i)$, $(z_a-z_i)$ and $v_i$ measurable, an available choice of $\bm{u_a}$ and $k_{\sss H}$ can always be found such that 
\begin{equation}
\left| {g\left( {z_a^k(u_a^k) \!-\! z_i^k,z_{i*}^k \!-\! z_i^k,v_a^k \!-\! v_i^k} \right)} \right| = \left| {\frac{{z_c^* \!-\! z_i^0}}{{{k_H}}}} \right| \!\le\! b,
\end{equation}
where $R_s\le\left\|z_a^k-z_i^k\right \|_2\le R_o$. 
The proof is completed. 
\end{proof}

\subsection{Proof of Theorem \ref{property}}\label{pf-th3}
\begin{proof}
We prove this theorem by contradiction. 

For i), suppose there are two distinct convex polygons $\mathcal{V}_{i,a}^{p}$ and $\mathcal{V}_{i,b}^{p}$ and the cardinal number $|\mathcal{V}_{i,a}^{p}| \ge |\mathcal{V}_{i,b}^{p}|$, then we obtain
\begin{equation}
\mathcal{V}_{i,\Delta}^{p}=\mathcal{V}_{i,a}^{p}\backslash (\mathcal{V}_{i,a}^{p}\cap\mathcal{V}_{i,b}^{p})\neq \emptyset.
\end{equation}
Then,  $\forall i'\in \mathcal{V}_{i,\Delta}^{p}$, $i'$ is not covered by $\mathcal{V}_{i,b}^{p}$, which renders a contradiction by Definition \ref{polygon}. 
Therefore, $\mathcal{V}_{i}^{p}$ is unique. 

For ii), the conclusion obviously holds when $|\mathcal{V}_{i}^{p}|=2$, thus we consider nontrivial cases where $|\mathcal{V}_{i}^{p}|\ge3$. 
First, when $|\mathcal{V}_{i}^{p}|=3$, there are only two vertex adjacent to $i$, here denoted as ${j_1}$ and ${j_2}$, and let $d_{ij_k}\!=\!\|z_{j_k}\!-\!z_{i}\|_2,k=1,2$.
In this case, $\mathcal{P}(z_{j_1},d_{ij_1}) \cap \mathcal{P}(z_{j_2},d_{ij_2}) = \emptyset$ if and only if the states $z_{i}$, $z_{j_2}$ and $z_{j_2}$ are linear dependent, which contradicts with convex vertex condition of $i$. 
Therefore, it follows that
\begin{equation} \label{2-neighbor}
\mathcal{P}_i(d_{ij_1},d_{ij_2})=\mathcal{P}(z_{j_1},d_{ij_1}) \cap \mathcal{P}(z_{j_2},d_{ij_2})\neq \emptyset.
\end{equation}

When $|\mathcal{V}_{i}^{p}|>3$, $\forall j_3 \in \left\{\mathcal{V}_{i}^{p}\backslash \{i\cup j_1\cup j_2\}\right\} $, utilizing (\ref{2-neighbor}), we easily obtain 
\begin{equation} \label{22-neighbor}
\mathcal{P}_i(d_{ij_1},d_{ij_3})\neq \emptyset,~\mathcal{P}_i(d_{ij_2},d_{ij_3})\neq \emptyset. 
\end{equation}
Then, what we need to do is to prove 
\begin{equation} \label{3-neighbor}
\mathcal{P}_i(d_{ij_1},d_{ij_3})\cap\mathcal{P}_i(d_{ij_2},d_{ij_3})\!\neq\! \emptyset. 
\end{equation}
Similarly, suppose $\mathcal{P}_i(d_{ij_1},d_{ij_3})\cap\mathcal{P}_i(d_{ij_2},d_{ij_3})\neq \emptyset$. 
This case is equivalent to the states of four vertex satisfying
\begin{equation}\label{sa}
  z_{j_3}-z_i=\alpha_1(z_{j_1}-z_i)+\alpha_2(z_{j_2}-z_i),
\end{equation}
where $\alpha_1\le0$ and $\alpha_2\le0$. 
Note that $\alpha_1=0$ (or $\alpha_2=0$) indicates $j_1$ (or $j_2$), $j_3$ and $i$ are on the same line, and $\alpha_1,\alpha_2$ are not zero at the time. 
Consequently, we have
\begin{equation}
{z_i} = \frac{{{z_{j_3}} - {\alpha _1z_{j_1}} - {\alpha _2z_{j_2}}}}{{1 - {\alpha _1} - {\alpha _2}}}, 
\end{equation}
which means $z_i$ is a convex combination of $z_{j_k}$ ($k=1,2,3$) and thus contradicts with convex vertex condition of $i$. 
Therefore, it follows that if $\mathcal{P}_i(d_{ij_1},d_{ij_3})\neq\! \emptyset$ and $\mathcal{P}_i(d_{ij_1},d_{ij_3})\neq\! \emptyset$, then $\mathcal{P}_i(d_{ij_2},d_{ij_3})\neq\! \emptyset$. 
By this transitivity property, we have 
\begin{equation}
\mathcal{Z}_i^{f_0}={ \mathop \bigcap \limits_{\mathclap{j \in \{\mathcal{V}_{i}^{p}\backslash i \}}} {\mathcal{P}}(z_j, d_{ij})}\neq \emptyset. 
\end{equation}
If there exists $j \!\in\! \{\mathcal{N}_i^{out}\backslash\mathcal{V}_{i}^{p}\}$, it is also covered by $\mathcal{V}_{i}^{p}$. 
Likewise, by the convex properties, $ {\mathcal{P}}(z_j, d_{ij}) \cap \mathcal{Z}_i^{f_0} =\emptyset$. 

To sum up, define the feasible position set as 
\begin{equation}
\mathcal{Z}_i^{f}={ \mathop \bigcap \limits_{\mathclap {j \in \mathcal{N}_i^{out} }} {\mathcal{P}}(z_j, d_{ij})}, 
\end{equation}
and by the definition of $\mathcal{P}(z, R)$, it yields that $\forall z \in \mathcal{Z}_{i}^{f}, {\left\| {z - z_j} \right\|_2} < {\left\| {{z_i} - z_j} \right\|_2}$. 
The proof is completed. 
\end{proof}

\subsection{Proof of Lemma \ref{lh44}}
\begin{proof}
For simplicity of expression, we directly consider the excitation is injected at the steady stage. 

Based on Theorem \ref{th01}, at steady stage, the error $(z(t)-{c}t\! \cdot \!\bm{1} -s)$ is extremely minor and negligible. 
Then, the global steady state evolution is represented as 
\begin{equation}
{z_{ss}} (t)= {q_1^e}{(p^e_1)^\mathsf{T}}u \cdot t +{q^e_1}{(p^e_1)^\mathsf{T}}z(0) +  \sum\limits_{i = 2}^N {\frac{1}{{{\lambda _i}}}{q^e_i}{(p^e_i)^\mathsf{T}}u}.  
\end{equation}
Note that the desired state deviation vector $h$ contained in $u$ only incurs a constant offset in $z_{ss}$. 
Therefore, ignore $h$ and let $u=[0,\cdots,0,u_e,0,\cdots,0,u_c]^\mathsf{T}$,  
and we obtain 
\begin{equation}\label{backwards}
\dot z_{ss} = ({p^e_{\sss 1j}}{u_e} + {p^e_{\sss 1N}{u_c}})\bm{1}. 
\end{equation}

When ${u_e}{u_c}>0$, it means that the excitation of $r_a$ aims to strengthen the movement in the direction of the original leader. 
In this case, arbitrary $u_e$ satisfying ${u_e}{u_c}>0$ is available to meet the indirect controllability. 

When ${u_e}{u_c}<0$, it means that the excitation of $r_a$ aims to strengthen the movement in the direction of ${u_e}$. 
Thus, one infers from (\ref{backwards}) that, to counteract the influence of the original leadership $u_c$ and ensure $\dot z_{i}u_e>0 $, the following condition must hold, given by 
\begin{equation}
|p^e_{\sss 1j}{u_e}| >|p^e_{\sss 1N}{u_c}|. 
\end{equation}
The proof is completed. 
\end{proof}

\subsection{Proof of Theorem \ref{th44}}
\begin{proof}
The core of this proof is to analyze the influence of the in-neighbors of $r_i$. 

First, we begin with a simple situation where $\forall j'\in \{\mathcal{N}_i^{in}\backslash{j} \}$, $a_{j'j}=0$, i.e., other in-neighbors of $r_i$ do not interact with $r_j$. 
Recalling the dynamics of $r_i$ given by (\ref{eq-1}), when $r_j$ is under the excitation of $u_e$, we have
\begin{align} \label{zhankai}
{\dot z}_i = &a_{ij}(z_j-{z_i}-h_j +h_i)+ \!\!\!\!\! \sum\limits_{j' \in \{ \mathcal{N}_i^{in}\backslash{j} \}} {\!\!\!\!\! a_{ij'}(z_{j'}- {z_i}-h_{j'} +h_i)  } \nonumber  \\
= &a_{ij}(u_et+z_j^0-{z_i}-h_j +h_i)  \nonumber\\
&+\!\!\!\!\! \sum\limits_{j' \in \{ \mathcal{N}_i^{in}\backslash{j} \}} {\!\!\!\!\!  a_{ij'}(u_ct+z_{j'}^0- {z_i}-h_{j'} +h_i)  }.
\end{align}
Let $b_{ij}=a_{ij}(z_{j'}^0-h_{j'} +h_i)$, $\bar{b}_{ij}=\!\!\!\!\!\sum\limits_{j' \in \{ \mathcal{N}_i^{in}\backslash{j} \}}\!\!\!\!\!a_{ij'}(z_{j'}^0-h_{j'} +h_i)$, and $\bar{a}_{ij}=\!\!\!\!\! \sum\limits_{j' \in \{ \mathcal{N}_i^{in}\backslash{j} \}}{\!\!\!\!\!  a_{ij'}}$. 
Then, (\ref{zhankai}) is rewritten as 
\begin{align} \label{weifen}
{\dot z}_i &= a_{ij}u_e t-a_{ij}z_i+b_{ij}+ \bar{a}_{ij}u_c t-\bar{a}_{ij}z_i+\bar{b}_{ij}  \nonumber  \\
&=(a_{ij}u_e+\bar{a}_{ij}u_c)t-(a_{ij}+\bar{a}_{ij})z_i+(b_{ij}+\bar{b}_{ij}) \nonumber \\
&=b_1 t-b_2 z_i + b_3. 
\end{align}
Note that (\ref{weifen}) is a first-order constant coefficient non-homogeneous linear equation, and leveraging constant variation method, the solution is given as 
\begin{equation}
{z_i}(t) = \frac{{{b_1}}}{{{b_2}}}t + (\frac{{{b_1}}}{{b_2^2}} - \frac{{{b_3}}}{{{b_2}}})({e^{ - {b_2}t}} - 1).
\end{equation}
Then, we obtain 
\begin{equation}
{\dot z_i}(t) = \frac{{{b_1}}}{{{b_2}}}=\frac{{  a_{ij}u_e+\bar{a}_{ij}u_c }}{{ a_{ij}+\bar{a}_{ij} }}. 
\end{equation}
The next step is the same as the proof of Lemma \ref{lh44}. 
It follows that whether $r_i$ is indirectly controllable is determined by $(z_c^*-z_i^0)u_c$ and $a_{ij}u_e+\bar{a}_{ij}u_c$. 
Accordingly, one infers that $r_i$ is indirect controllable if and only if (\ref{str}) is satisfied. 

Finally, consider $\exists j'\in \{\mathcal{N}_i^{in}\backslash{j} \}$, $a_{j'j}>0$, i.e., other in-neighbors of $r_i$ are also influenced by $r_j$. 
Then, (\ref{zhankai}) is rewritten as 
\begin{align} \label{zhankai2}
{\dot z}_i \!= \!a_{ij}(z_j\!-\!{z_i}\!-\!h_j\! +\!h_i)\!+ \!\!\!\!\!\!\! \sum\limits_{j' \in \{ \mathcal{N}_i^{in}\backslash{j} \}} \!\!\!{\!\!\!\!\! a_{ij'}(z_{j'}(z_j)\!-\! {z_i}\!-\!h_{j'} \!+\!h_i)  },
\end{align}
where $z_{j'}(z_j)$ is also determined by $z_j$ and no longer linearly increasing. 
However, $a_{j'j}>0$ yields $|{z_{j'}}(t) - {z_j}(t)| < |{u_c}t + z_{j'}^0 - {z_j}(t)|$, which incurs $z_i(t)$ more closer to $z_c^*$ compared with that when $a_{j'j}=0$ at the same time. 
Therefore, by (\ref{str}), it is sufficient to guarantee the indirect controllability of $r_i$. 
The proof is completed. 
\end{proof}

\begin{IEEEbiographynophoto}{Yushan Li}
(S'19) received the B.E. degree in School of Artificial Intelligence and Automation from Huazhong University of Science and Technology, Wuhan, China, in 2018. 
He is currently working toward the Ph.D. degree with the Dept. of Automation, Shanghai Jiaotong University, Shanghai, China. 
He is a member of Intelligent of Wireless Networking and Cooperative Control group. 
His research interests include robotics, security of cyber-physical system, and distributed computation and optimization in multi-agent networks. 
\end{IEEEbiographynophoto}

\begin{IEEEbiographynophoto}{Jianping He} 
(SM'19) is currently an associate professor in the Department of Automation at Shanghai Jiao Tong University. He received the Ph.D. degree in control science and engineering from Zhejiang University, Hangzhou, China, in 2013, and had been a research fellow in the Department of Electrical and Computer Engineering at University of Victoria, Canada, from Dec. 2013 to Mar. 2017. His research interests mainly include the distributed learning, control and optimization, security and privacy in network systems.

Dr. He serves as an Associate Editor for IEEE Open Journal of Vehicular Technology and KSII Trans. Internet and Information Systems. He was also a Guest Editor of IEEE TAC, International Journal of Robust and Nonlinear Control, etc. He was the winner of Outstanding Thesis Award, Chinese Association of Automation, 2015. He received the best paper award from IEEE WCSP'17, the best conference paper award from IEEE PESGM'17, and was a finalist for the best student paper award from IEEE ICCA'17.
\end{IEEEbiographynophoto}
\vspace{-20pt}

\begin{IEEEbiographynophoto}{Xuda Ding}
(S'20) received the B.E. degree and M.E. degree in School of Automation from North China Electric Power University, China, in 2016 and 2019, respectively. 
He is currently working toward the Ph.D. degree with the Dept. of Automation, Shanghai Jiaotong University, Shanghai, China. 
He is a member of Intelligent of Wireless Networking and Cooperative Control group. 
His research interests include robotics, learning theory and its application in practical systems. 
\end{IEEEbiographynophoto}
\vspace{-20pt}

\begin{IEEEbiographynophoto}{Cai Lin}
(F'20) received the M.A.Sc. and Ph.D. degrees (awarded Outstanding Achievement in Graduate Studies) in electrical and computer engineering from the University of Waterloo, Waterloo, Canada, in 2002 and 2005, respectively. Since 2005, she has been with the Department of Electrical and 
Computer Engineering at the University of Victoria, where she is currently a Professor. 
She is an NSERC E.W.R. Steacie Memorial Fellow. Her research interests span several areas in communications and networking, with a focus on network protocol and architecture design supporting emerging multimedia traffic and the Internet of Things. 

She was a recipient of the NSERC Discovery Accelerator Supplement (DAS) Grants in 2010 and 2015, respectively, and the best paper awards of IEEE ICC 2008 and IEEE WCNC 2011. She has co-founded and chaired the IEEE Victoria Section Vehicular Technology and Communications Joint Societies Chapter. 
She has been elected to serve the IEEE Vehicular Technology Society Board of Governors, 2019 - 2021. She has served as an Area Editor for IEEE TRANSACTIONS ON VEHICULAR TECHNOLOGY, a member of the Steering Committee of the IEEE TRANSACTIONS ON BIG DATA (TBD) and IEEE TRANSACTIONS ON CLOUD COMPUTING (TCC), an Associate Editor of the IEEE INTERNET OF THINGS JOURNAL, IEEE TRANSACTIONS ON WIRELESS COMMUNICATIONS, IEEE TRANSACTIONS ON VEHICULAR TECHNOLOGY, IEEE TRANSACTIONS ON COMMUNICATIONS, EURASIP Journal on Wireless Communications and Networking, International Journal of Sensor Networks, and Journal of Communications and Networks (JCN), and as the Distinguished Lecturer of the IEEE VTS Society. She has served as a TPC co-chair for IEEE VTC2020-Fall, and a TPC symposium co-chair for IEEE Globecom’10 and Globecom’13. She is a Registered Professional Engineer in British Columbia, Canada.
\end{IEEEbiographynophoto}
\vspace{-20pt}

\begin{IEEEbiographynophoto}{Xinping Guan}
(F'18) received the B.S. degree in Mathematics from Harbin Normal University, Harbin, China, in 1986, and the Ph.D. degree in Control Science and Engineering from Harbin Institute of Technology, Harbin, China, in 1999. He is currently a Chair Professor with Shanghai Jiao Tong University, Shanghai, China, where he is the Dean of School of Electronic, Information and Electrical Engineering, and the Director of the Key Laboratory of Systems Control and Information Processing, Ministry of Education of China. 
Before that, he was the Professor and Dean of Electrical Engineering, Yanshan University, Qinhuangdao, China. 

Dr. Guan's current research interests include industrial cyber-physical systems, wireless networking and applications in smart factory, and underwater networks. He has authored and/or coauthored 5 research monographs, more than 270 papers in IEEE Transactions and other peer-reviewed journals, and numerous conference papers. 
As a Principal Investigator, he has finished/been working on many national key projects. He is the leader of the prestigious Innovative Research Team of the National Natural Science Foundation of China (NSFC). 
Dr. Guan received the First Prize of Natural Science Award from the Ministry of Education of China in both 2006 and 2016, and the Second Prize of the National Natural Science Award of China in both 2008 and 2018. 
He was a recipient of IEEE Transactions on Fuzzy Systems Outstanding Paper Award in 2008. He is a National Outstanding Youth honored by NSF of China, Changjiang Scholar by the Ministry of Education of China and State-level Scholar of New Century Bai Qianwan Talent Program of China.
\end{IEEEbiographynophoto}

\end{document}